 \documentclass[final,3p,times,twocolumn]{elsarticle}
\usepackage{amssymb}
\biboptions{sort&compress}

\journal{NIM B362(2015)116}

\begin{document}

\begin{frontmatter}



\title{New data on cross-sections of deuteron induced nuclear reactions on gold up to 50 MeV and comparison of production routes of medically relevant Au and Hg radioisotopes}


\author[1]{F. T\'ark\'anyi}
\author[2]{A. Hermanne}
\author[1]{F. Ditr\'oi\corref{*}}
\author[1]{S. Tak\'acs}
\author[2]{R. Adam-Rebeles}
%\author[3]{M. Baba}
%\author[4]{B.M.A. Mohsena}
\author[3]{A.V. Ignatyuk}
\cortext[*]{Corresponding author: ditroi@atomki.hu}

\address[1]{Institute for Nuclear Research, Hungarian Academy of Sciences (ATOMKI),  Debrecen, Hungary}
\address[2]{Cyclotron Laboratory, Vrije Universiteit Brussel (VUB), Brussels, Belgium}
%\address[3]{Cyclotron Radioisotope Center (CYRIC), Tohoku University, Sendai, Japan}
%\address[4]{Nuclear Research Center – Egyptian Atomic Energy Authority, Cairo, Egypt}
\address[3]{Institute of Physics and Power Engineering (IPPE), Obninsk, Russia}

\begin{abstract}
Investigations of cross-sections of deuteron induced nuclear reactions on gold were extended up to 50 MeV by using the standard stacked foil irradiation technique and high resolution gamma-ray spectrometry.  New cross-sections  are reported for the $^{197}$Au(d,xn)$^{197m,197g,195m,195g,193m,193g}$Hg and $^{197}$Au(d,x)$^{198m,198g,196m,196g,195,194}$Au nuclear reactions. The application for production of the medically relevant isotopes $^{198}$Au and $^{195m,195g,197m,197g}$Hg is discussed, including the comparison with other charged particle induced production routes. The possible use of the $^{197}$Au(d,x)$^{197m,197g,195m,193m}$Hg and $^{196m,196g}$Au reactions for  monitoring deuteron beam parameters is also investigated.
\end{abstract}


\end{frontmatter}

% \linenumbers

\section{Introduction}
\label{1}
The main motivation for the present measurement was to extend and complete the database for activation cross-sections of deuteron induced nuclear reactions on metals (gold in this case) for accelerator technology in the frame of the IAEA Coordinated Research Project on Fusion Evaluated Nuclear Data Library (FENDL-3 \cite{1}).
In the literature numerous works, dealing with deuteron activation of Au, were found \cite{2,3,4,5,6,7,8,9,10,11,12,13,14,15,16,17,18}, including our previous work up to 40 MeV deuteron energy \cite{19}. This is probably due to the simple target preparation of monoisotopic stable Au and the broad range of practical applications. We present here new experimental deuteron activation cross-sections up to 50 MeV, obtained as secondary results from studies of Ga-Ni targets electro-deposited on gold backings. Some proton induced reactions on gold were already proposed as beam monitor reactions \cite{20}. We have further investigated and evaluated the available data for monitoring deuteron beam parameters (energy, intensity) using gold targets.

\section{Experimental and data evaluation}
\label{2}
The experimental data were obtained from a 50 MeV deuteron irradiation at the CGR930 cyclotron in Louvain la Neuve (LLN). The details of the experimental technique are described in the work of Adam-Rebeles et al.: “$^{68}$Ge/$^{68}$Ga production revisited: Excitation curves, target preparation and chemical separation – purification” \cite{21}. The targets were Ga-Ni alloy (70-30\%) deposited with the electroplating technique described in section 3 of \cite{21} on Au (thickness 25 $\mu$m for deuteron experiment) or Cu (12.5 $\mu$m, for proton experiment) foils as backing material. 
In the stack irradiated with deuterons, twenty two Ga-Ni targets on Au backing were interleaved with 55 Al (98 $\mu$m and 49.6 $\mu$m) and 15 Ni (23.9 $\mu$m) monitor/degrader foils. The nominal thickness of the Ga-Ni layer was 17.3 $\mu$m. The targets were mounted in Faraday cup like target holders assembled with a long collimator. The beam current was kept constant at about 100 nA (± 5 \%) during the 60 minute irradiation. Because the Au backing got the bombardment from the side of the Ga-Ni layer, a loss of Hg isotopes during the irradiation could not occur.
The produced activities were assessed using an HPGe detector coupled with the acquisition/analysis software GENIE 2000 (CANBERRA, USA). The samples were measured several times in order to follow the decay. The source to detector distance was varied from 45 cm, shortly after the end of bombardment, down to 5 cm for the last series of measurements in order to keep dead-times below 10\%. The HPGe detector efficiency curves were determined in the experimental geometry with a standard calibrated $^{152}$Eu source. The activities of the irradiated samples were measured without chemical separation.
The initial beam parameters were estimated from the accelerator settings and from current beam measurements. Final data were obtained by the simultaneous re-measurement of the cross sections of the $^{27}$Al(d,x)$^{22,24}$Na and comparison with the recommended data in the IAEA-TECDOC \cite{22} allowing to introduce corrections where needed.  The agreement of experimental cross sections of the monitor reactions (adjusted beam current and primary beam energy) with the recommended data is shown in Fig. 1.
The decay data of the investigated radionuclides and the Q-values of the contributing formation reactions were taken from the on-line NUDAT2 data base \cite{23} and the BNL Q-value calculator \cite{24} (see Table 1). 
The uncertainty on each cross-section was estimated \cite{25} by taking the square root of the sum of all individual contributions in quadrature: absolute abundance of the γ-ray taken from NUDAT2 (5\%), determination of the peak areas including statistical errors (2–8\%), the number of target nuclei including non-uniformity (5\%), detector efficiency (5\%) and incident particle intensity (5\%). The total uncertainty of the cross-section values was evaluated to approximately 10 to 12\%.
The uncertainty of the median beam energy on the first foil was estimated to be about ±0.3 MeV. Due to possible variations and inhomogeneity in thickness (5\%), energy spread and straggling, the uncertainties for the energy is increasing throughout the stack and reaches a maximum of  $\pm$ 1.2  MeV in the last  foil.

\begin{figure}
\includegraphics[scale=0.3]{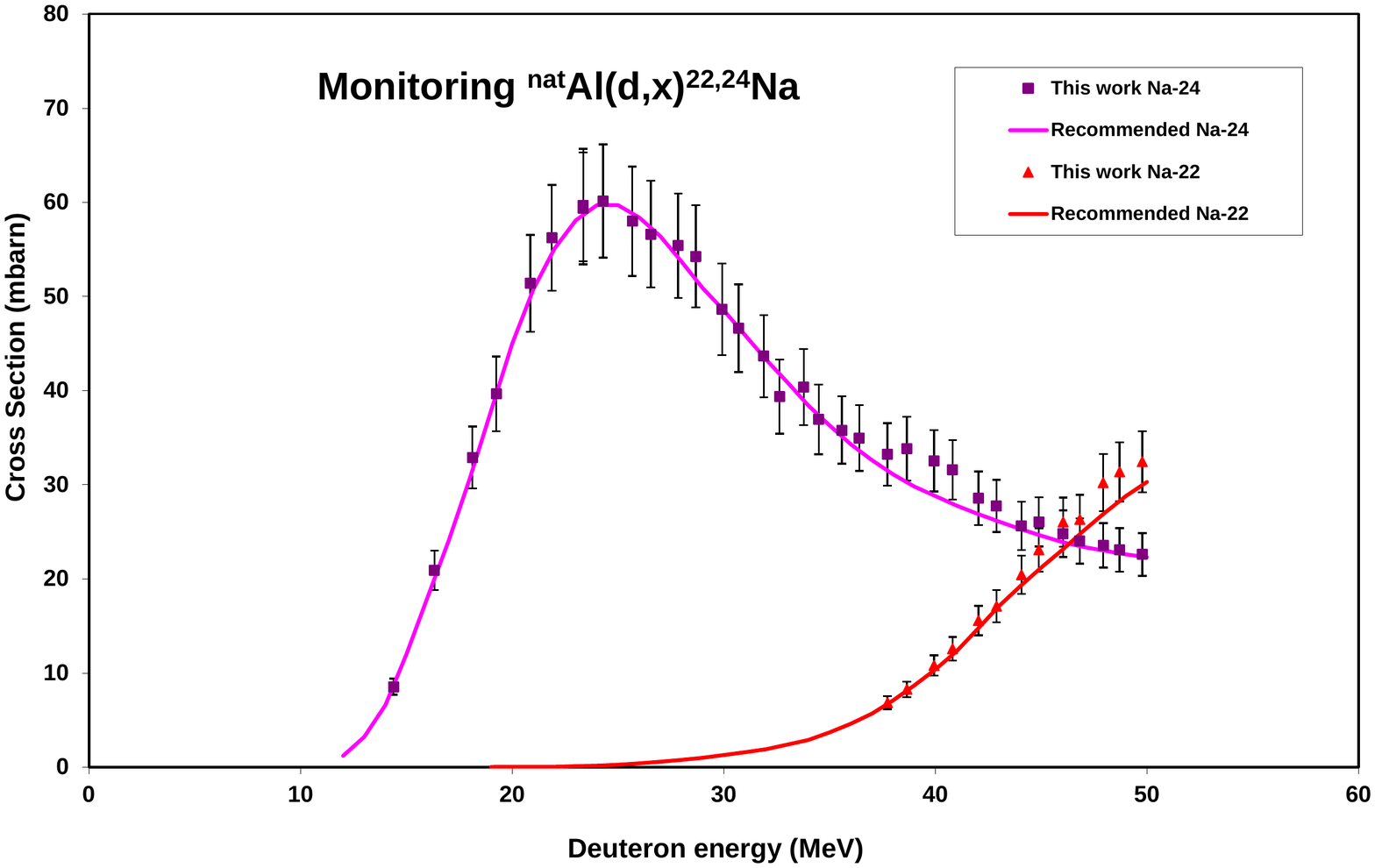}
\caption{Monitoring of the deuteron beam parameters with the $^{nat}$Al(d,x)$^{24,22}$Na reactions and comparison with the recommended values}
\end{figure}

\begin{table*}[t]
\tiny
\caption{Decay and nuclear characteristics of the investigated reaction products, contributing reactions and their Q-values \cite{23, 24}.}
\centering
\begin{center}
\begin{tabular}{|p{0.4in}|p{0.3in}|p{0.4in}|p{0.3in}|p{0.3in}|p{0.5in}|p{0.5in}|}
\hline
\textbf{Nuclide\newline Spin\newline Isomeric level} & \textbf{Half-life} & \textbf{
Decay\newline method} & \textbf{E$_{\gamma}$(keV)} & \textbf{I$_{\gamma}$(\%)
} & \textbf{Contributing process} & \textbf{Q-value(keV)} \\
\hline
$^{197m}$Hg\newline 13/2$^{+}$\newline 298.93 keV & 23.8 h & EC 8.6 \%\newline IT 91.4\% & 
133.98\newline 279.01 & 33.5\newline 6 & $^{197}$Au(d,2n) & -3607.03 \\
\hline
$^{197g}$Hg\newline 1/2$^{-}$ & 64.14 h & EC 100 \% & 77.351\newline 191.437 & 
18.7\newline 0.632 & $^{197}$Au(d,2n)\newline $^{197m}$Hg decay & -3607.03 \\
\hline
$^{195m}$Hg\newline 13/2$^{+}$\newline 176.07 keV & 41.6 h & EC 45.8 \%\newline IT 54.2\% & 
261.75\newline 387.87\newline 560.27 & 31\newline 2.18\newline 7.1 & $^{197}$Au(d,4n) & -19290.6 \\
\hline
$^{195g}$Hg\newline 1/2$^{-}$ & 10.53 h & EC 100 \% & 
180.11\newline 207.1\newline 261.75\newline 585.13\newline 599.66\newline 779.80\newline 1111.04\newline 1172.38 & 
1.95\newline 1.6\newline 1.6\newline 2.04\newline 1.83\newline 7.0\newline 1.48\newline 1.28 & $^{197}$Au(d,4n)\newline $^{195m}$Hg decay 
& -19290.6 \\
\hline
$^{193m}$Hg\newline 13/2$^{(+)}$\newline 140.765 & 11.8 h & IT 7.2 \%\newline EC 92 \% & 
257.99\newline 407.63\newline 573.26\newline 932.57 & 49\newline 32\newline 26\newline 12.5 & $^{197}$Au(d,6n) & -35370.3 
\\
\hline
$^{193g}$Hg\newline 3/2$^{-}$ & 3.8 h & EC 100 \% & 
186.56\newline 381.60\newline 861.11\newline 1118.84 & 15.2\newline 16\newline 12.4\newline 8.0 & $^{197}$Au(d,6n)\newline $^{195m
}$Hg decay & -35370.3 \\
\hline
$^{198m}$Au\newline 12$^{-}$\newline 811.715 keV & 2.272 d & IT 100\% & 
97.21\newline 180.31\newline 204.10\newline 214.89\newline 333.82 & 69\newline 49\newline 39\newline 77.3\newline 18 & $^{197}$Au(d,p) & 
4287.764 \\
\hline
$^{198g}$Au2$^{-}$ & 2.6947 d & $\beta^{-}$: 100 \% & 411.80205 & 
95.62 & $^{197}$Au(d,p) & 4287.764 \\
\hline
$^{196m2}$Au\newline 12$^{-}$\newline 595.664 keV & 9.6 h & IT 100\% & 
137.69\newline 147.81\newline 168.37\newline 188.27\newline 285.49\newline 316.19 & 1.3\newline 43.5\newline 7.8\newline 30.0\newline 4.4\newline 3.0 & $^{197}$
Au(d,p2n) & -10296.96 \\
\hline
$^{196g}$Au2$^{-}$ & 6.1669 d & EC 93.0 \%\newline $\beta^{-}$: 7.0 \% & 
333.03\newline 355.73\newline 426.10 & 22.9\newline 87\newline 6.6 & $^{197}$Au(d,p2n)\newline $^{196m}$Au 
decay & -10296.96 \\
\hline
$^{195g}$Au\newline 3/2$^{+}$ & 186.09 d & EC 100 \% & 98.88\newline 129.757 & 
11.2\newline 0.842 & $^{197}$Au(d,p3n)\newline $^{195m}$Au\newline decay$^{195}$Hg 
decay & -16938.27 \\
\hline
$^{194}$Au\newline 1$^{-}$ & 38.02 h & EC 100 \% & 293.548\newline 328.464\newline1468.882 & 
10.58\newline 60.4\newline 6.61 & $^{197}$Au(d,p4n)\newline $^{194}$Hg decay & -25317.5 \\
\hline
\end{tabular}

\end{center}
\begin{flushleft}
\tiny{\noindent The Q-values shown in Table 1 refer to the formation of the ground state. Increase absolute Q-values for isomeric states with level energy of the isomer. When complex particles are emitted instead of individual protons and neutrons the absolute Q-values have to be decreased by the respective binding energies of the compound particles: np-d, +2.2 MeV; 2np-t, +8.48 MeV; 2p2n-$\alpha$, +28.30 MeV}
\end{flushleft}

\end{table*}

%\setcounter{table}{1}

%\begin{table*}[t]
%\tiny
%\caption{continued}
%\centering
%\begin{center}

%\end{center}
%\end{table*} 

\section{Nuclear model calculations}
\label{3}
The cross-sections of the investigated reactions calculated using the pre-compound model codes ALICE-IPPE-D \cite{26} and EMPIRE-II-D \cite{27} were reported already in our previous work covering the energy range up to 40 MeV. Independent data for isomers with ALICE-D code was obtained by using the isomeric ratios calculated with EMPIRE-D. The experimental data are also compared with the curves given in three versions of the TENDL library, TENDL 2009  \cite{28} used in our previous publications, TENDL-2013 \cite{28, 29}  and TENDL 2014 \cite{30} relying on newer versions of the TALYS code family. The aim is to show the importance of experimental data and to demonstrate the improvements in code results for part of the reactions. The TENDL library is based on both default and adjusted TALYS (version 1.6) calculations \cite{31}.

\section{Cross sections}
\label{4}
Activation cross-sections for the $^{197}$Au(d,xn)$^{197m,197g,195m,195g}$Hg and $^{197}$Au(d,x)$^{198g,196m,196g,195,194}$Au nuclear reactions were measured. The new experimental cross section data are shown in Figs. 2-11 in comparison with literature values and with the predictions of theoretical codes. The results from our previous work \cite{19}, are also presented in the original form, i.e. the different measurement are separately as series 1 (ser-1) and series 2 (ser-2). The results from the present work are also presented in the same way. In the cases of some isotopes, because of the availability of more precise nuclear data, a correction was necessary, which is indicated in the legends. It is also indicated in the legends, if cumulative cross section is given (total). The numerical data are collected in Table 2 and 3. The cross section values of mercury radionuclides are due to direct production via (d,xn) reactions. The gold radio-products are produced directly via (d,pxn) reactions and/or additionally through the decay of the shorter-lived isobaric parent mercury radioisotope (cum). The ground state of the produced radioisotopes can, apart from the direct production, also be produced through the internal transition of an isomeric state. The cross section is marked with (m+), when the half-life of the isomeric state is significantly shorter than to the half-life of the ground state and the cross section for ground state formation was deduced from spectra obtained after nearly complete decay of the parent isomeric state.  
It was impossible to deduce production cross sections for the Pt radioisotopes, due to the low cross sections and the complex spectra on the Ga-Ni/Au target.
Detailed comparisons with the earlier experimental data and with the theoretical calculations up to 40 MeV were discussed in our previous work \cite{19}. The new data are in good agreement with our earlier results in the overlapping energy range and properly follow the trend above 40 MeV.

\subsection{Production of radioisotopes of mercury}
\label{4.1}

\subsubsection{Cross sections for the $^{197}$Au(d,x)$^{197m}$Hg reaction}
\label{4.1.1}

The 23.8 h isomeric state decays for 91.4 \% to the 64.1 h half-life ground state and for the rest to short-lived $^{197m}$Au that decays to stable $^{197g}$Au. The production cross sections of the $^{197m}$Hg are shown in Fig 2, together with the earlier experimental results and theoretical estimations. Five earlier investigations were found for this product; by Krisanfov et al. up to 6 MeV \cite{18}, by Long et al. \cite{9} and Wenrong et al. \cite{8} cross section data up to 14 MeV, by Vandenbosch et al. up to 22 MeV \cite{11} and by Chevarier et al.  up to 32 MeV \cite{3}. The agreement with the former experimental results is acceptable. All theoretical approaches give the proper shape of the excitation function, but there is an energy shift in ALICE-D towards the lower energies. The codes EMPIRE-D and the new TENDL-2014 (identical to TENDL-2013) strongly overestimate, only the older TENDL-2009 results give an acceptable approach except around the maximum. 

\begin{figure}
\includegraphics[scale=0.3]{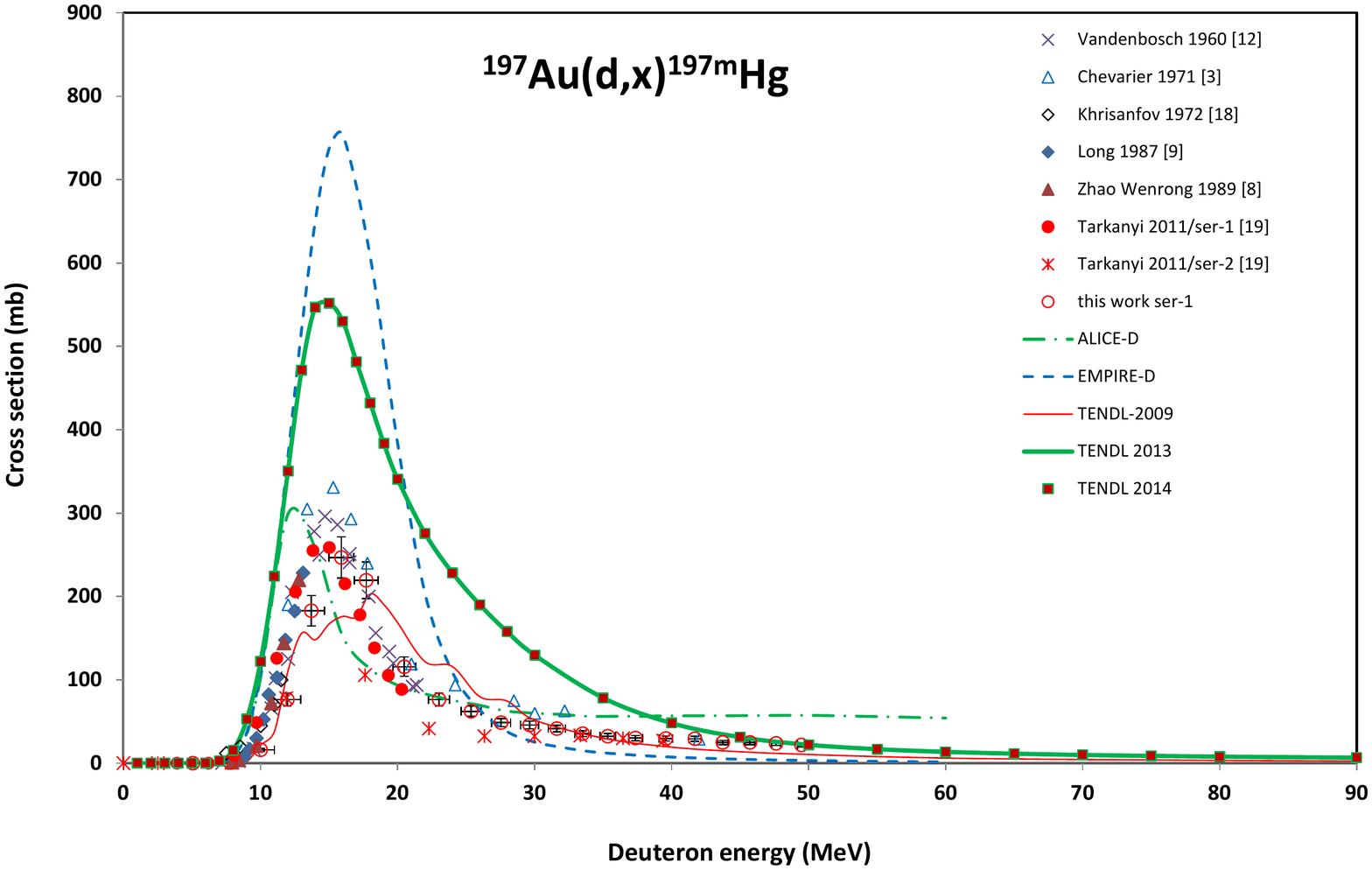}
\caption{Experimental and theoretical cross sections for the formation of $^{197m}$Hg by the deuteron bombardment of gold}
\end{figure}

\subsubsection{Cross sections for the $^{197}$Au(d,x)$^{197g}$Hg(m+) reaction}
\label{4.1.2}

The cross-section data for direct production of the ground state were obtained after correction for contribution of decay from the isomeric state. As both states have independent $\gamma$-lines and as the half-lives are somewhat different, analytical separation of the ingrowth is possible. In Fig. 3 we reproduce all available experimental data for the direct production of $^{197g}$Hg together with the theoretical results. The situation with the theoretical model codes is similar to the previous case, except that here the newest TENDL versions (2014 identical to 2013) give acceptable agreement, while the older TENDL-2009 gives strange and unacceptable results. The agreement with the previous experimental results is acceptable, except for the values of Zhao Wenrong et al. \cite{8} and Long et al. \cite{9} around the maximum.

\begin{figure}
\includegraphics[scale=0.3]{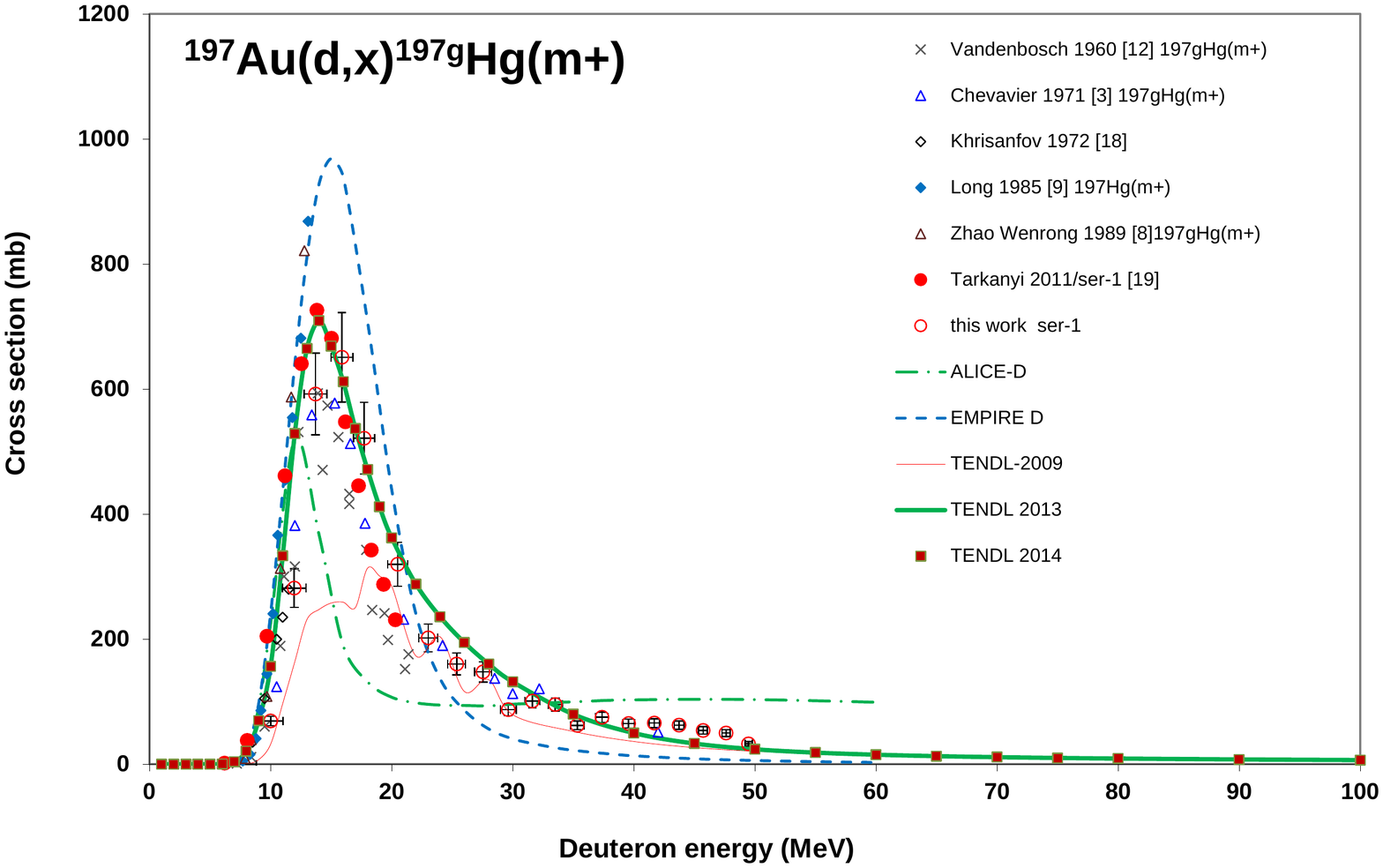}
\caption{Experimental and theoretical cross sections for the formation of $^{197g}$Hg(m+) by the deuteron bombardment of gold}
\end{figure}

\subsubsection{Cross sections for the $^{197}$Au(d,x)$^{195m}$Hg reaction}
\label{4.1.3}

The radionuclide 195Hg has two long-lived states: a longer-lived high spin isomer ($^{195m}$Hg, T$_{1/2}$ = 41.6 h, I$^\pi$ = 13/2$^+$) and the shorter-lived ground state $^{195g}$Hg (T$_{1/2}$ = 10.53 h, I$^\pi$ = 1/2$^-$). We obtained the production cross section of the 41.6 hour half-life, high spin isomeric state as it can be seen in Fig.4. The agreement with the former experimental results is acceptable except the maximum values of Chevarier et al. \cite{3}. In this case the theoretical model calculations show large disagreement with each other and with the experimental values. The improvement of the new TENDL-2014 (identical to 2013) is clearly seen: those two give the best agreement with our new experimental results below 40 MeV.  

\begin{figure}
\includegraphics[scale=0.3]{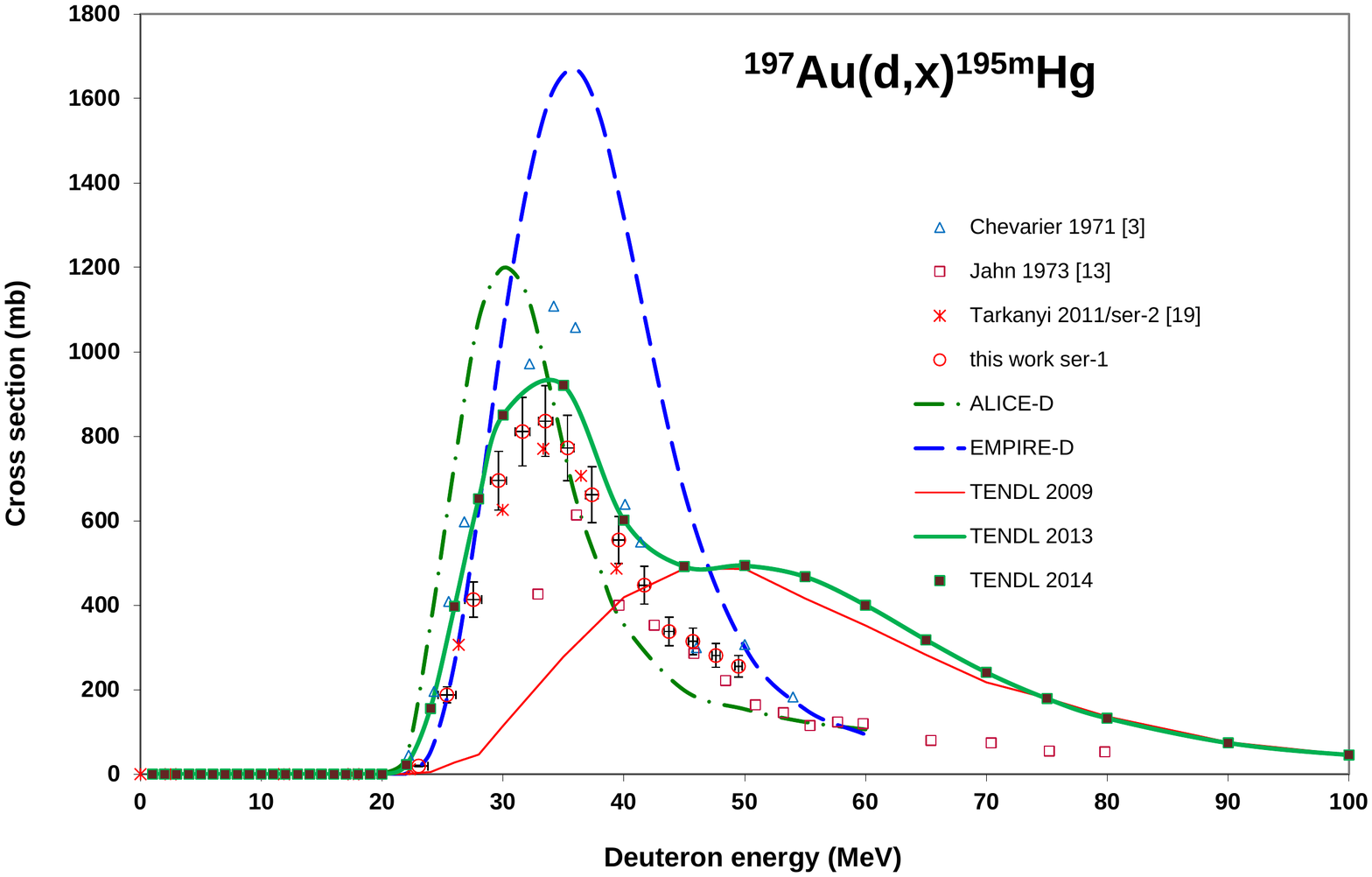}
\caption{Experimental and theoretical cross sections for the formation of $^{195m}$Hg by the deuteron bombardment of gold}
\end{figure}

\subsubsection{Cross sections for the $^{197}$Au(d,x)$^{195g}$Hg reaction}
\label{4.1.4}
The independent cross section for the direct ground state formation can be determined if a correction is made for ingrowth by the partial isomeric decay of $^{195m}$Hg (T$_{1/2}$ = 41.6 h, 54.2 \% IT). The agreement in Fig. 5 with the former experiments is acceptable except the data of Jahn et al. \cite{13}. In this case only the EMIRE-D gives a somewhat good approach, except for the maximum values where it also fails. The new TENDL calculations improve only the shape but not the values.

\begin{figure}
\includegraphics[scale=0.3]{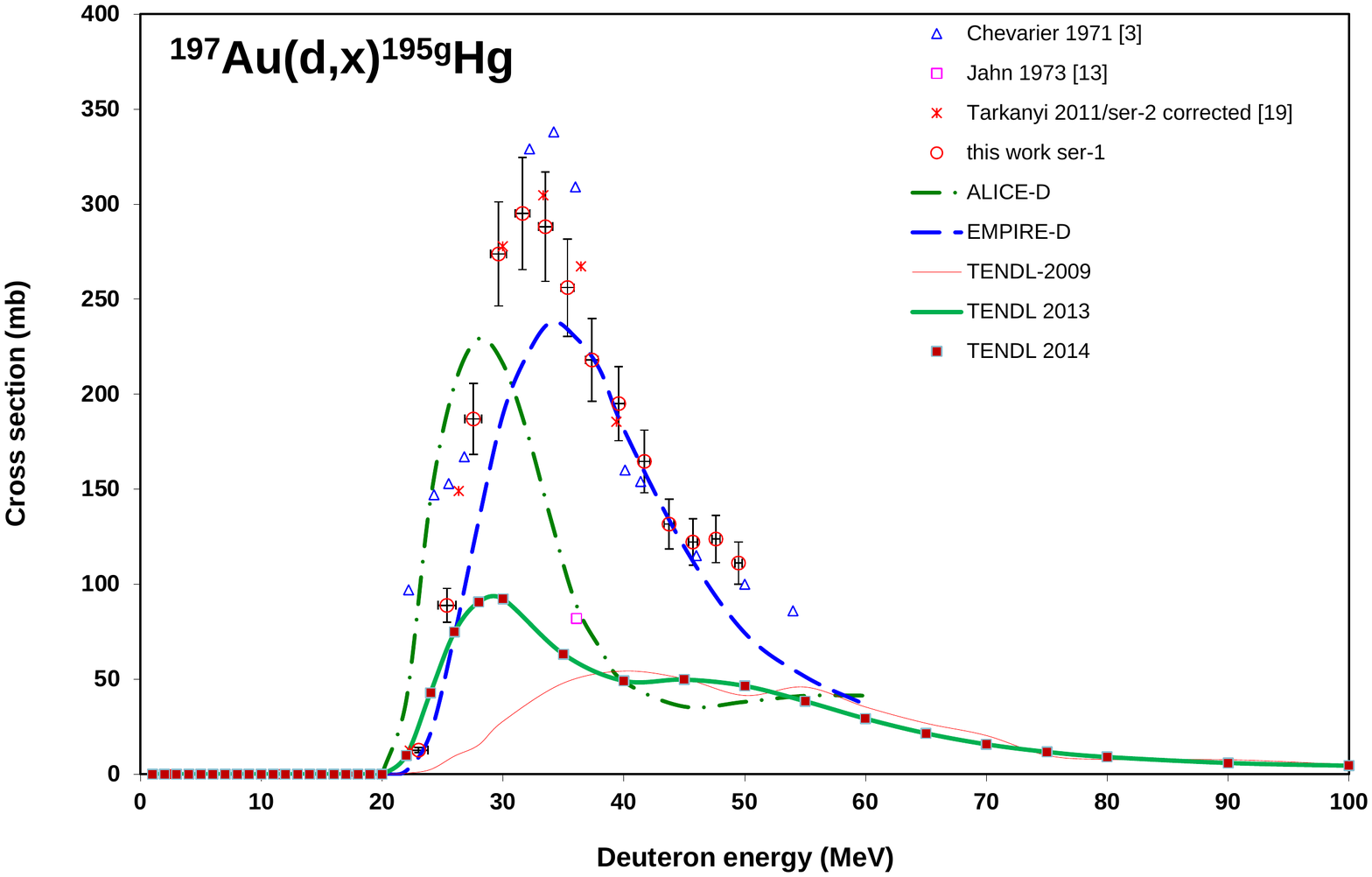}
\caption{Experimental and theoretical cross sections for the formation of $^{195g}$Hg by the deuteron bombardment of gold}
\end{figure}

\subsubsection{Cross sections for the $^{197}$Au(d,x)$^{193m}$Hg reaction}
\label{4.1.5}
The radionuclide $^{193}$Hg has two long-lived states: a longer-lived, high spin isomer ($^{193m}$Hg, T$_{1/2}$ = 11.8h, I$^\pi$ = 13/2$^+$, IT: 7.2 \%, EC: 92 \%and the shorter-lived ground state $^{193g}$Hg (T$_{1/2}$ = 3.80 h, I$^\pi$ = 3/2$^-$). The measured excitation function is shown in Fig. 6. The new data are in good agreement with the previous experimental data of Jahn et al. \cite{13}. The improvement in TENDL from 2009 to 2014 (identical to 2013) is clearly seen but the maximum value is still largely underestimated.

\begin{figure}
\includegraphics[scale=0.3]{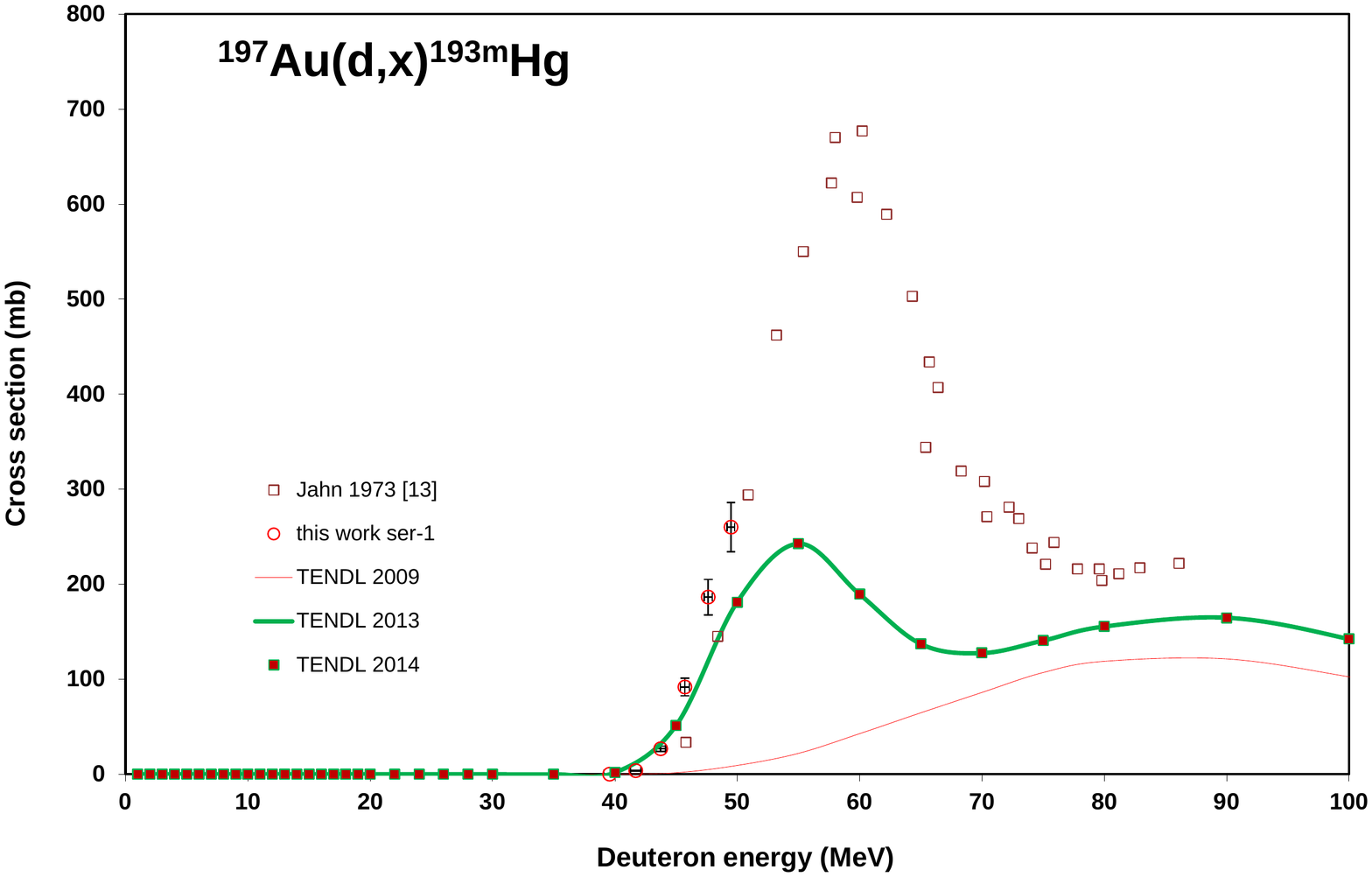}
\caption{Experimental and theoretical cross sections for the formation of $^{193m}$Hg by the deuteron bombardment of gold}
\end{figure}

\subsubsection{Cross sections for the $^{197}$Au(d,x)$^{193g}$Hg reaction}
\label{4.1.6}
The cross sections for direct production of the ground state of $^{193g}$Hg were obtained by subtracting the contribution from the decay of the isomeric state.  No earlier experimental data were found. The comparison with the prediction of the TENDL-2014 (identical to TENDL-2013) library shows good agreement in the studied energy domain (Fig 7.)

\begin{figure}
\includegraphics[scale=0.3]{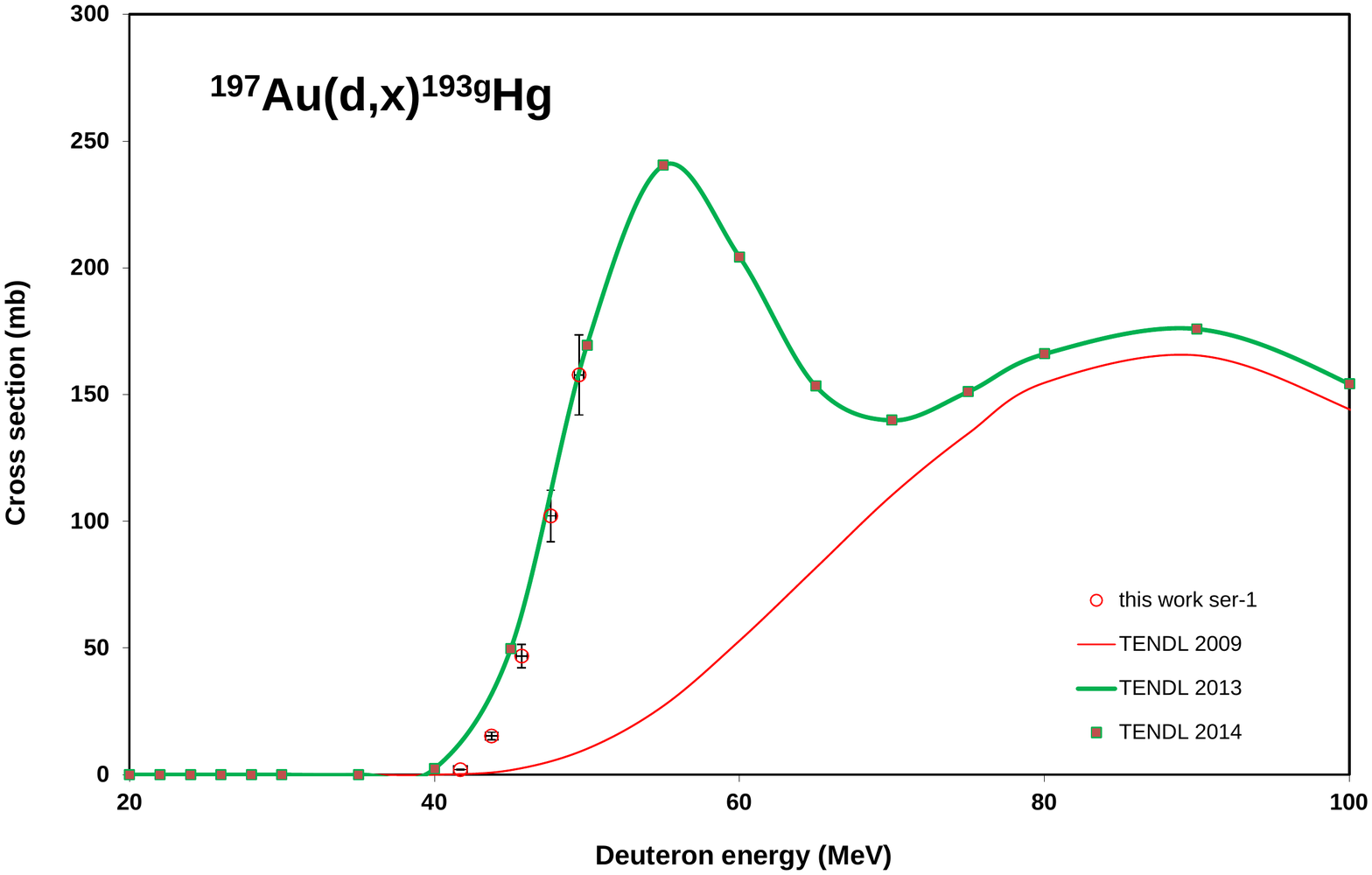}
\caption{Experimental and theoretical cross sections for the formation of $^{193g}$Hg by the deuteron bombardment of gold}
\end{figure}

\subsection{Production of radioisotopes of gold}
\label{4.2}

\subsubsection{Cross sections for the $^{197}$Au(d,x)$^{198m}$Au reaction}
\label{4.2.1}
We could measure the cross section data for formation of both longer-lived states, i.e. the ground state and the isomeric state. The cross sections of this very high spin isomeric state $^{198m}$Au (T$_{1/2}$ = 2.272 d, I$^\pi$ = 12$^-$, 811.715 keV, IT: 100 \%) are shown in Fig. 8 in comparison with our earlier data up to 40 MeV and with the theoretical results. In all TENDL calculations unfortunately only the total productions cross sections for $^{198}$Au are presented, but the change between the three different versions is enormous. The ALICE-D and EMPIRE-D results are acceptable.

\begin{figure}
\includegraphics[scale=0.3]{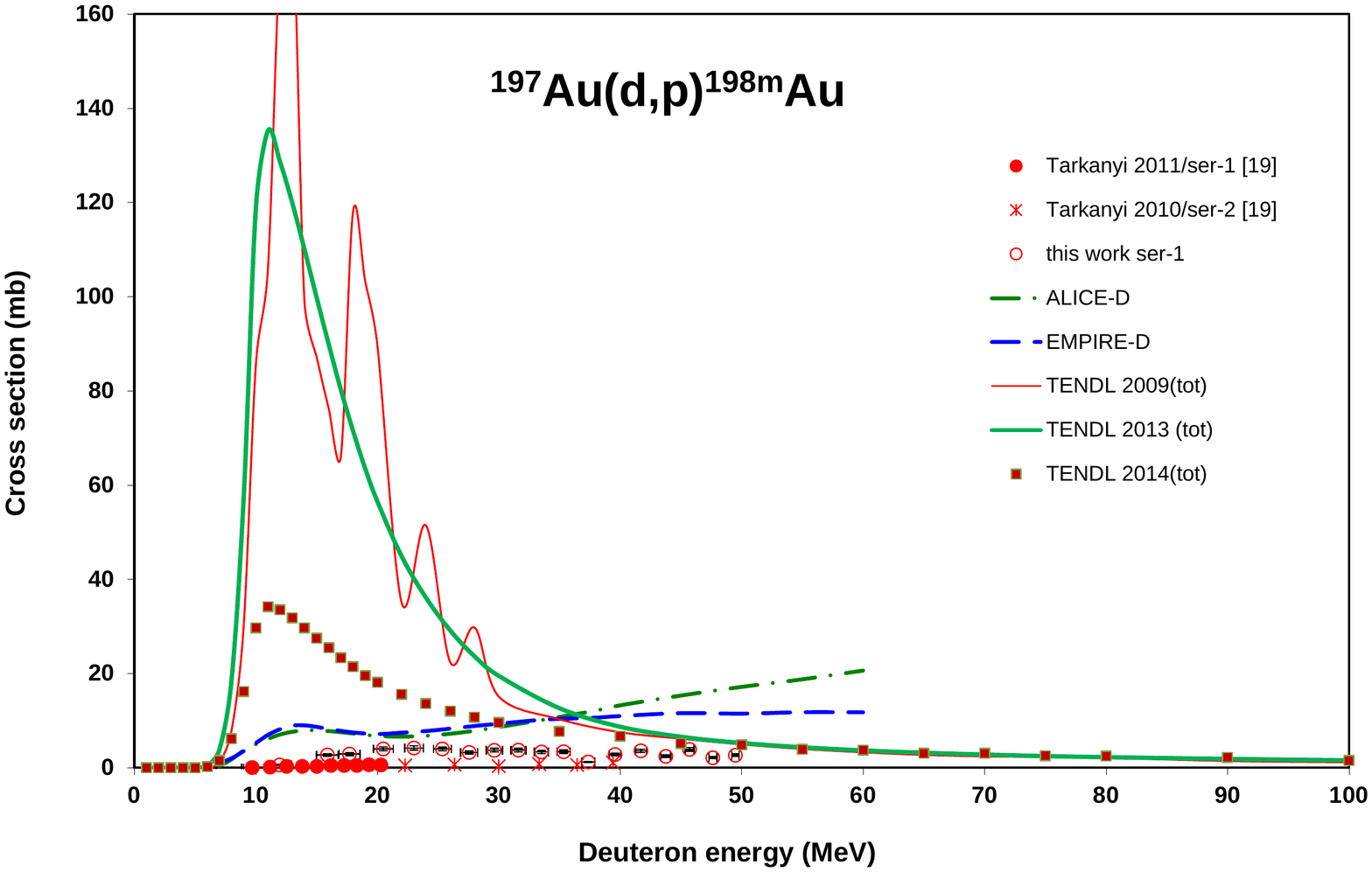}
\caption{Experimental and theoretical cross sections for the formation of $^{198m}$Au by the deuteron bombardment of gold}
\end{figure}

\subsubsection{Cross sections for the $^{197}$Au(d,x)$^{198g}$Au reaction}
\label{4.2.2}
The experimental and theoretical results for direct production of the ground state $^{198g}$Au (T$_{1/2}$ = 2.6947 d, I$^\pi$ = 2$^-$), after correction for contribution of 100\% isomeric transitions from $^{198m}$Au, are shown in Fig. 9.  Seven datasets were found in the literature and are all in rather good agreement in the overlapping energy intervals although near the maximum the values of  Nassif et al. \cite{14} are low while those of Wenrong et al. \cite{8} and  Sandoval \cite{15} are too high compared to the average trend.
ALICE-D and EMPIRE-D describe well the experimental behavior in shape with a 10-15\% underestimation of the maximum value. While the TENDL-2009 (total) has a non-physical shape, the predictions for TENDL-2013 (total) are approaching within 60\% the experimental values. The newest TENDL-2014 is worse and too low by a factor of 5.

\begin{figure}
\includegraphics[scale=0.3]{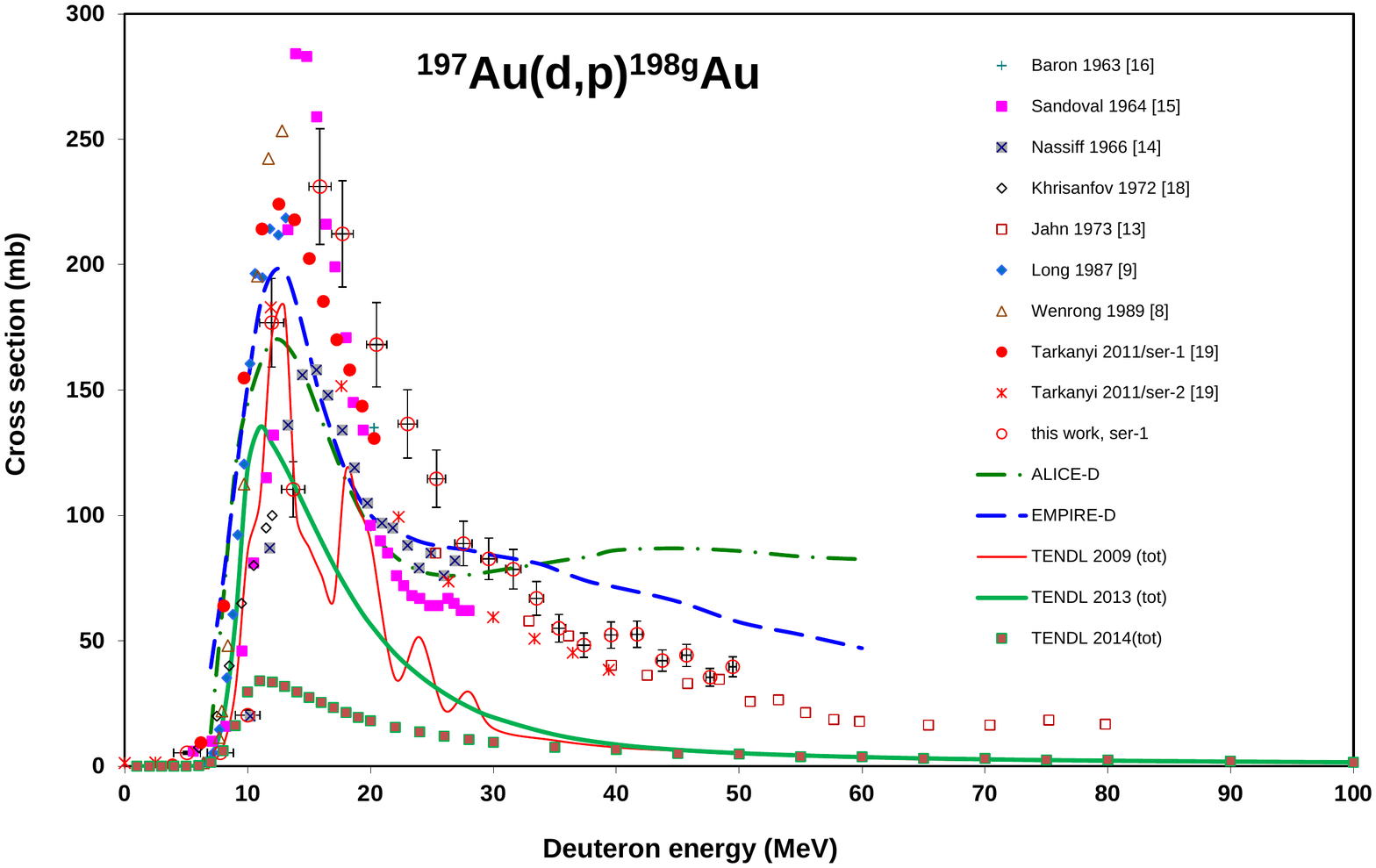}
\caption{Experimental and theoretical cross sections for the formation of $^{198g}$Au by the deuteron bombardment of gold}
\end{figure}

\subsubsection{Cross sections for the $^{197}$Au(d,x)$^{196m2}$Au reaction}
\label{4.2.3}
The cross section data for formation of the high spin isomeric state $^{196m2}$Au (T$_{1/2}$ = 9.6 h, I$^\pi$ = 12$^-$) are shown in Fig.10. Good agreement with the earlier experimental data of Chevarier et al. \cite{3} and Jahn  et al. \cite{13} is noticed . The ALICE and EMPIRE calculations strongly underestimate the experimental values. There are no data in the TENDL libraries for the isomeric states separately and hence they are not represented here. $^{196m1}$Au (T$_{1/2}$ = 8.1 s) could not be measured in this experiment.

\begin{figure}
\includegraphics[scale=0.3]{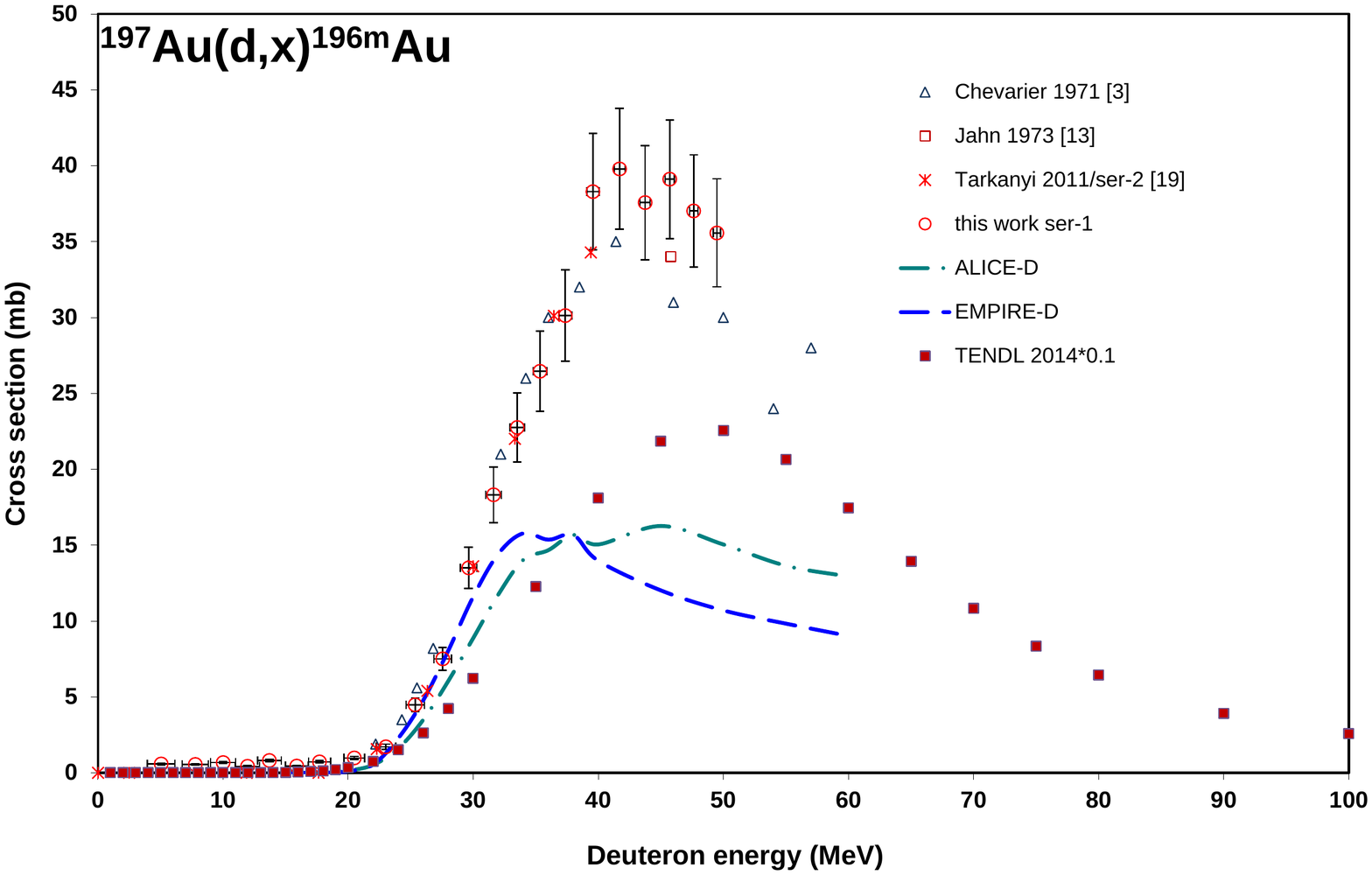}
\caption{Experimental and theoretical cross sections for the formation of $^{196m}$Au by the deuteron bombardment of gold}
\end{figure}

\subsubsection{Cross sections for the $^{197}$Au(d,x)$^{196g}$Au reaction}
\label{4.2.4}
The experimental results for direct production of the $^{196g}$Au ground state (T$_{1/2}$ = 6.1669 d, I$^\pi$ = 2$^-$), after correction for contribution of 100\% isomeric transitions from $^{196m2}$Au, but including the total decay of the direct formation of the 8.1 sec $^{196m1}$Au (I$^\pi$ = 5$^+$), are shown in Fig. 11. The agreement with the former experimental data is acceptable. The ALICE and EMPIRE calculations strongly underestimate the experimental values. The data in TENDL libraries for the total formation of $^{196}$Au overestimate the results for the direct formation. 

\begin{figure}
\includegraphics[scale=0.3]{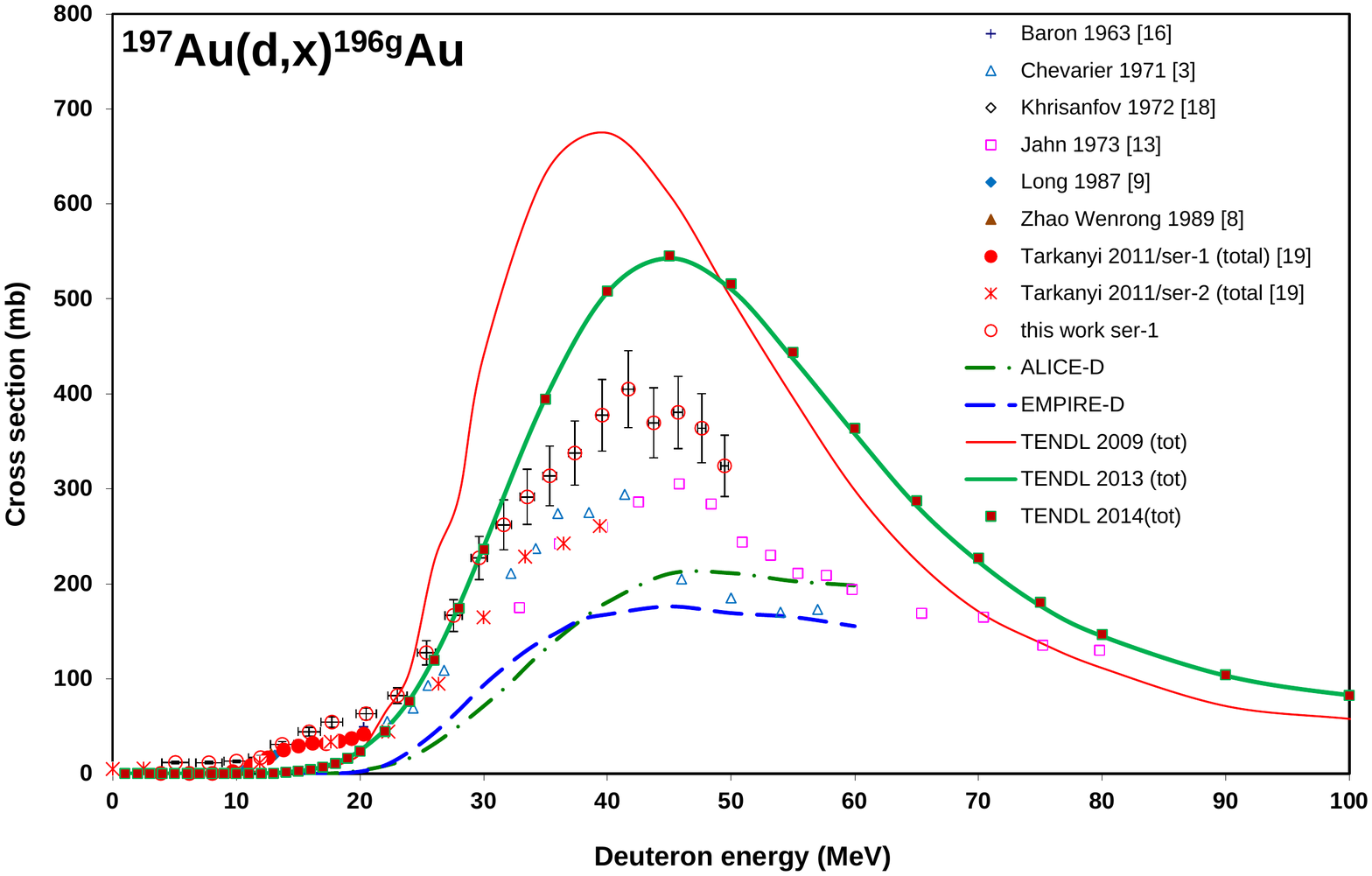}
\caption{Experimental and theoretical cross sections for the formation of $^{196g}$Au by the deuteron bombardment of gold}
\end{figure}

\subsubsection{Cross sections for the $^{197}$Au(d,x)$^{195}$Au reaction}
\label{4.2.5}
The measured excitation function for production of $^{195}$Au (T$_{1/2}$ = 186.098 d, I$^\pi$ = 3/2$^+$) contains the decay of short-lived $^{195m}$Au (T$_{1/2}$ = 30.5 s, IT 100\%) and the complete decay of parents $^{195m}$Hg(T$_{1/2}$ = 41.6 h) and $^{195g}$Hg (T$_{1/2}$ = 10.53 h). No earlier experimental data were found in the literature. The comparison of our data with the different TENDL versions is shown in Fig. 12. TENDL-2014 (identical to 2013) shows a large improvement over the earlier 2009 version and acceptable agreement under 40 MeV.

\begin{figure}
\includegraphics[scale=0.3]{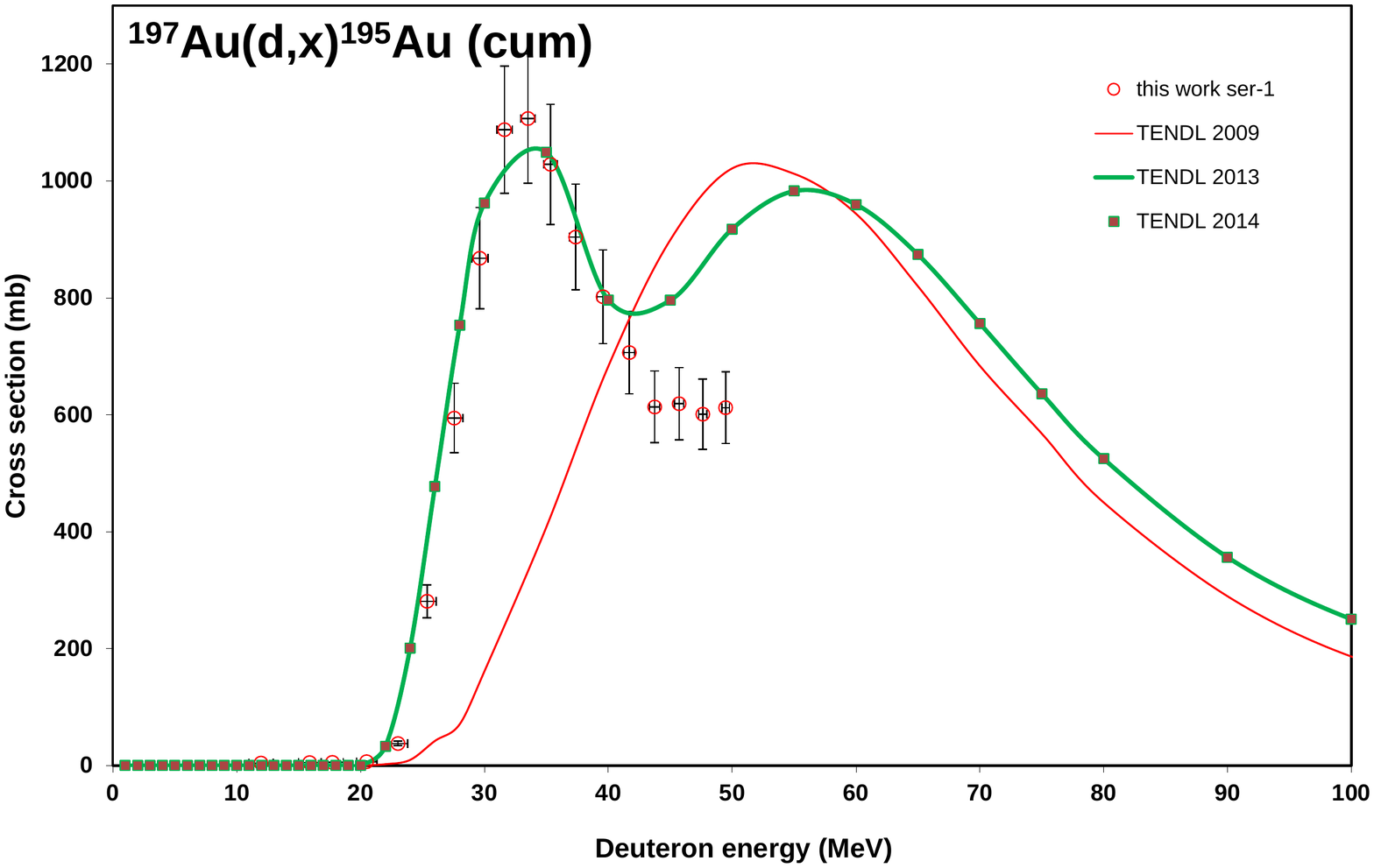}
\caption{Experimental and theoretical cross sections for the formation of $^{195}$Au by the deuteron bombardment of gold}
\end{figure}

\subsubsection{Cross sections for the $^{197}$Au(d,x)$^{194}$Au reaction}
\label{4.2.6}
The excitation function for production of $^{194}$Au (T$_{1/2}$ = 38.02 h, I$^\pi$ = 1$^-$) is shown in Fig. 13 in comparison with the experimental datasets from the literature and with the theory. The agreement with the existing experimental data is acceptable, except with the data of Jahn et al \cite{13}, which are lower than ours above 45 MeV. Both the older and the new TENDL versions give acceptable agreement in the investigated energy range and confirm the Jahn et al. values up to 60 MeV. Also the predictions of ALICE-D and EMPIRE-D represent well the experimental data.

\begin{figure}
\includegraphics[scale=0.3]{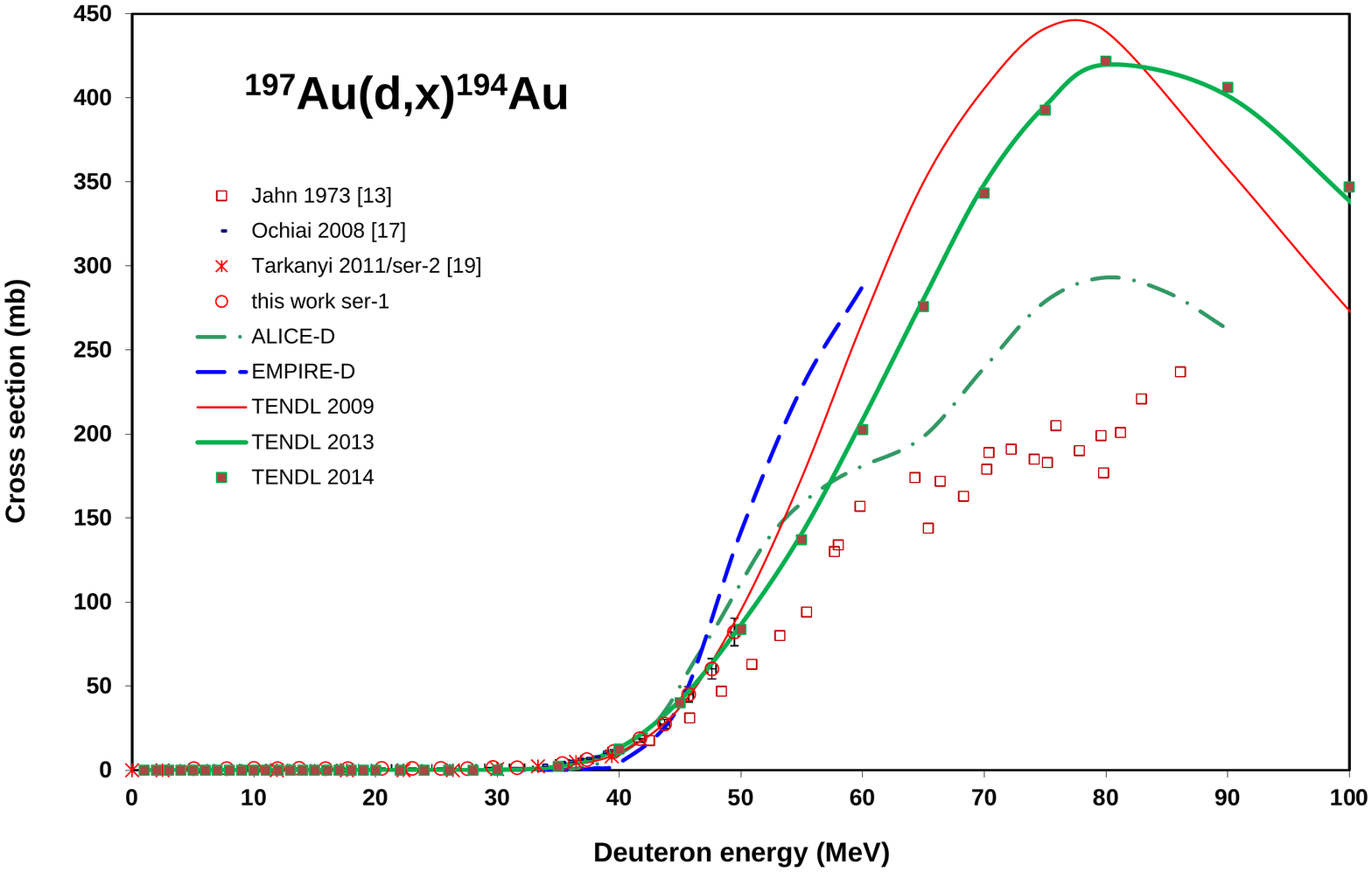}
\caption{Experimental and theoretical cross sections for the formation of $^{194}$Au by the deuteron bombardment of gold}
\end{figure}

\subsubsection{Cross sections for the $^{197}$Au(d,x)$^{193,192}$Au reactions}
\label{4.2.7}
The gamma lines of $^{193}$Au (T$_{1/2}$=17.65 h) and $^{192}$Au (T$_{1/2}$= 4.94 h) were also detected in our spectra, but they are produced mostly via their similar half-life parent $^{193}$Hg (T$_{1/2}$ = 3.80 h and 11.8 h) and $^{192}$Hg (T$_{1/2}$ = 4.85 h) radioisotopes.  On the basis of our spectra the direct production cross section can be deduced only with high uncertainties and are not reported in this paper.

\begin{table*}[t]
\tiny
\caption{Experimental cross sections of $^{197}$Au(d,xn)$^{197m,197g,195m,195g,193m,193g}$Hg nuclear reactions}
\centering
\begin{center}
\begin{tabular}{|r|r|r|r|r|r|r|r|r|r|r|r|r|r|}
\hline
\multicolumn{2}{|c|}{\textbf{E $\pm \Delta$E}} & \multicolumn{2}{|c|}{\textbf{$^{197m}$Hg}} & \multicolumn{2}{|c|}{\textbf{$^{197g}$Hg}} & \multicolumn{2}{|c|}{\textbf{$^{195m}$Hg}} & \multicolumn{2}{|c|}{\textbf{$^{195g}$Hg}}
 & \multicolumn{2}{|c|}{\textbf{$^{193m}$Hg}} & \multicolumn{2}{|c|}{\textbf{$^{193g}$Hg}} \\
\cline{3-14}
\multicolumn{2}{|c|}{\textbf{MeV}} & \multicolumn{12}{|c|}{\textbf{$\sigma \pm \Delta\sigma$ (mbarn)}} \\
\hline
5.1 & 1.1 & 0.22 & 0.02 &  & &  & &  & &  & &  & \\
\hline
7.8 & 1.1 & 1.0 & 0.1 &  & &  & &  & &  & &  & \\
\hline
10.0 & 1.0 & 16 & 2 & 69 & 7 &  & &  & &  & &  & \\
\hline
12.0 & 1.0 & 76 & 8 & 282 & 28 &  & &  & &  & &  & \\
\hline
13.7 & 0.9 & 183 & 18 & 592 & 59 &  & &  & &  & &  & \\
\hline
15.9 & 0.9 & 247 & 25 & 651 & 65 &  & &  & &  & &  & \\
\hline
17.7 & 0.9 & 220 & 22 & 521 & 52 &  & &  & &  & &  & \\
\hline
20.5 & 0.8 & 116 & 12 & 320 & 32 &  & &  & &  & &  & \\
\hline
23.0 & 0.8 & 76 & 8 & 202 & 20 & 19 & 2 & 13 & 1.3 &  & &  & \\
\hline
25.4 & 0.7 & 62 & 6 & 160 & 16 & 188 & 19 & 89 & 9 &  & &  & \\
\hline
27.6 & 0.7 & 49 & 5 & 148 & 15 & 414 & 41 & 187 & 19 &  & &  & \\
\hline
29.6 & 0.7 & 46 & 5 & 87 & 9 & 695 & 70 & 274 & 27 &  & &  & \\
\hline
31.6 & 0.6 & 42 & 4 & 101 & 10 & 811 & 81 & 295 & 30 &  & &  & \\
\hline
33.5 & 0.6 & 36 & 4 & 95 & 10 & 836 & 84 & 288 & 29 &  & &  & \\
\hline
35.3 & 0.6 & 33 & 3 & 62 & 6 & 773 & 77 & 256 & 26 &  & &  & \\
\hline
37.4 & 0.5 & 30 & 3 & 75 & 8 & 662 & 66 & 218 & 22 &  & &  & \\
\hline
39.6 & 0.5 & 30 & 3 & 65 & 7 & 555 & 56 & 195 & 20 & 0.2 & 0.02 &  & \\
\hline
41.7 & 0.4 & 30 & 3 & 66 & 7 & 448 & 45 & 165 & 17 & 4 & 0.4 & 2.1 & 0.2 
\\
\hline
43.7 & 0.4 & 25 & 3 & 63 & 6 & 338 & 34 & 132 & 13 & 27 & 3 & 7.1 & 0.7 
\\
\hline
45.7 & 0.4 & 24.9 & 3 & 54 & 5 & 315 & 32 & 122 & 12 & 92 & 9 & 24 & 2.4 
\\
\hline
47.6 & 0.3 & 24 & 2 & 50 & 5 & 281 & 28 & 124 & 12 & 186 & 19 & 42 & 4 
\\
\hline
49.5 & 0.3 & 22 & 2 & 33 & 3 & 256 & 26 & 111 & 11 & 260 & 26 & 60 & 6 
\\
\hline
\end{tabular}
\end{center}
\end{table*} 

\begin{table*}[t]
\tiny
\caption{Experimental cross sections of $^{197}$Au(d,x)$^{198m,198g,196m,196g,195,194}$Au}
\centering
\begin{center}
\begin{tabular}{|r|r|r|r|r|r|r|r|r|r|r|r|r|r|}
\hline
\multicolumn{2}{|c|}{\textbf{E $\pm \Delta$E}} & \multicolumn{2}{|c|}{\textbf{$^{198m}$Au}} & \multicolumn{2}{|c|}{\textbf{$^{198g}$Au}} & \multicolumn{2}{|c|}{\textbf{$^{196m}$Au}} & \multicolumn{2}{|c|}{\textbf{$^{196g}$Au}}
 & \multicolumn{2}{|c|}{\textbf{$^{195}$Au}} & \multicolumn{2}{|c|}{\textbf{$^{194}$Au}} \\
\cline{3-14}
\multicolumn{2}{|c|}{\textbf{MeV}} & \multicolumn{12}{|c|}{\textbf{$\sigma \pm \Delta\sigma$ (mbarn)}} \\
\hline
5.1 & 1.1 &  & & 5.4 & 0.5 & 0.6 & 0.1 & 11.8 & 1.2 &  & & 0.93 & 0.09 
\\
\hline
7.8 & 1.1 &  & & 5.4 & 0.5 & 0.6 & 0.1 & 11.7 & 1.2 &  & & 0.93 & 0.09 
\\
\hline
10.0 & 1.0 &  & & 20.4 & 2.0 & 0.7 & 0.1 & 13.2 & 1.3 &  & & 1.1 & 
0.11 \\
\hline
12.0 & 1.0 & 0.6 & 0.06 & 177 & 18 & 0.4 & 0.04 & 17 & 1.7 & 4.7 & 0.5 & 
0.99 & 0.10 \\
\hline
13.7 & 0.9 &  & & 110 & 11 & 0.8 & 0.1 & 31 & 3 &  & & 1.09 & 0.11 \\
\hline
15.9 & 0.9 & 2.7 & 0.3 & 231 & 23 & 0.5 & 0.03 & 44 & 5 & 5.1 & 0.5 & 
0.94 & 0.09 \\
\hline
17.7 & 0.9 & 2.9 & 0.3 & 212 & 21 & 0.7 & 0.1 & 54 & 6 & 5.8 & 0.6 & 
0.93 & 0.09 \\
\hline
20.5 & 0.8 & 4 & 0.4 & 168 & 17 & 1 & 0.1 & 63 & 6 & 6.9 & 0.7 & 1.15 & 
0.12 \\
\hline
23.0 & 0.8 & 4.2 & 0.4 & 137 & 14 & 2 & 0.2 & 82 & 8 & 38 & 4 & 1.14 & 
0.11 \\
\hline
25.4 & 0.7 & 4 & 0.4 & 115 & 12 & 5 & 0.4 & 127 & 13 & 281 & 28 & 1.07 & 
0.11 \\
\hline
27.6 & 0.7 & 3.2 & 0.3 & 89 & 9 & 8 & 0.8 & 167 & 17 & 595 & 60 & 0.94 & 
0.09 \\
\hline
29.6 & 0.7 & 3.7 & 0.4 & 83 & 8 & 14 & 1.4 & 227 & 23 & 868 & 87 & 1.6 & 
0.2 \\
\hline
31.6 & 0.6 & 3.7 & 0.4 & 79 & 8 & 18 & 2 & 262 & 26 & 1088 & 109 & 1.5 & 
0.2 \\
\hline
33.5 & 0.6 & 3.3 & 0.3 & 70 & 7 & 23 & 2 & 292 & 29 & 1107 & 111 &  & 
\\
\hline
35.3 & 0.6 & 3.4 & 0.3 & 55 & 6 & 27 & 3 & 314 & 31 & 1028 & 103 & 4 & 
0.4 \\
\hline
37.4 & 0.5 & 1.2 & 0.1 & 48 & 5 & 30 & 3 & 338 & 34 & 904 & 90 & 6.3 & 
0.6 \\
\hline
39.6 & 0.5 & 2.8 & 0.3 & 52 & 5 & 38 & 4 & 377 & 38 & 802 & 80 & 11.2 & 
1.1 \\
\hline
41.7 & 0.4 & 3.6 & 0.4 & 53 & 5 & 40 & 4 & 405 & 41 & 707 & 71 & 18.7 & 
2 \\
\hline
43.7 & 0.4 & 2.4 & 0.2 & 42 & 4 & 38 & 4 & 370 & 37 & 614 & 61 & 27 & 3 
\\
\hline
45.7 & 0.4 & 3.9 & 0.4 & 44 & 4 & 39 & 4 & 380 & 38 & 619 & 62 & 45 & 5 
\\
\hline
47.6 & 0.3 & 2.1 & 0.2 & 36 & 4 & 37 & 4 & 364 & 36 & 601 & 60 & 60 & 6 
\\
\hline
49.5 & 0.3 & 2.6 & 0.3 & 40 & 4 & 36 & 4 & 324 & 32 & 612 & 61 & 82 & 8 
\\
\hline
\end{tabular}
\end{center}
\end{table*} 

\section{Thick target yields}
\label{5}
The integral yields, calculated from spline fits to our experimental excitation functions, are shown in Figs. 14 and 15. The integral yields represent so called physical yields, i.e. activity at instantaneous production rates at 1 $\mu$A beam current \cite{32}. Only for a few activation products data are available from Dmitriev \cite{11}. The agreement for the $^{198g}$Au results is good, for $^{197g}$Hg and $^{196g}$Au we obtained different values.

\begin{figure}
\includegraphics[scale=0.3]{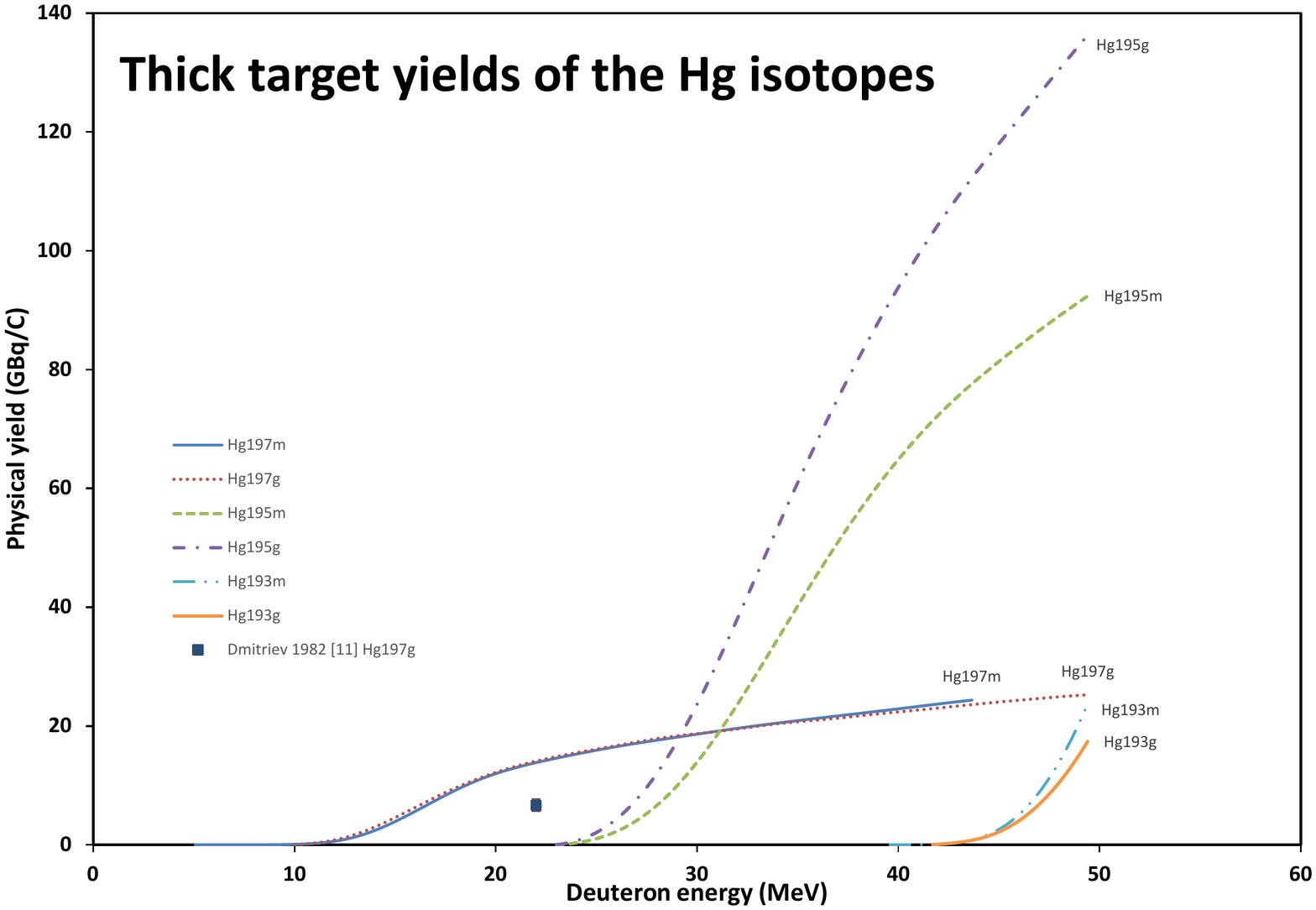}
\caption{Integral thick target yields for the formation mercury radioisotopes in deuteron  induced nuclear reaction on $^{197}$Au as a function of the energy}
\end{figure}

\begin{figure}
\includegraphics[scale=0.3]{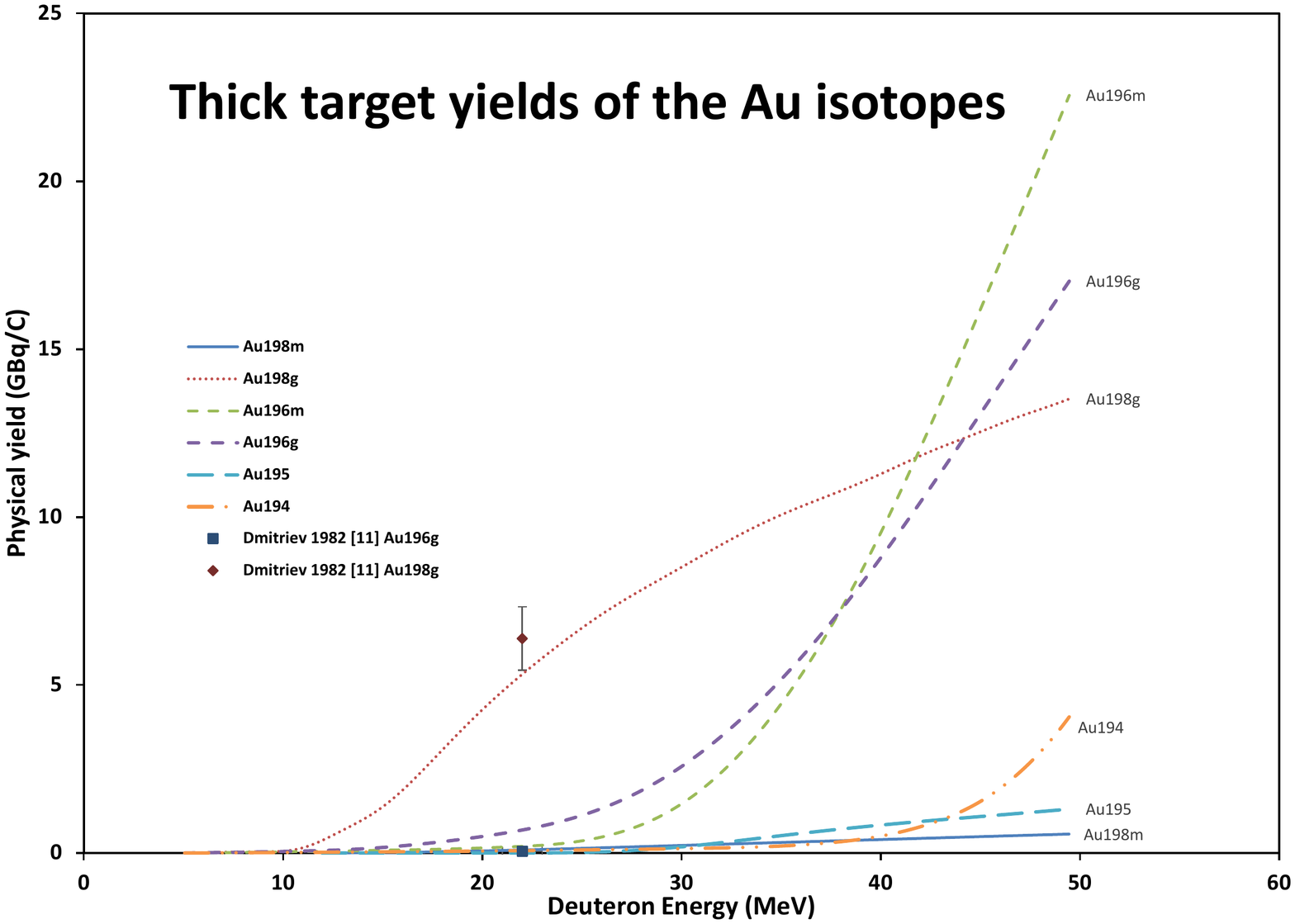}
\caption{Integral thick target yields for the formation gold radioisotopes in deuteron  induced nuclear reaction on $^{197}$Au as a function of the energy}
\end{figure}

\section{Application of nuclear data for production of medical radioisotopes}
\label{6}

Among the investigated the reaction products, the radionuclides $^{195m}$Au, $^{195g}$Au, $^{198}$Au, $^{197m}$Hg and $^{197g}$Hg are of interest in nuclear medicine. All of them could be used in radiotherapy \cite{33}, some of them also for diagnostic applications. These radionuclides can be produced by a large variety of nuclear reactions. In some of our earlier publications we investigated the production of these radioisotopes, using different accelerated particles and different targets \cite{19, 20, 34,35,36,37,38,29,40}. We review here the most important charged particle induced production routes, from the nuclear data point of view.

\subsection{$^{195m}$Au}
\label{6.1}
Several studies for the use of very short-lived $^{195m}$Au (T$_{1/2}$=30.5 s, E$_\gamma$  =262 keV) in nuclear angiocardiography were done in eighties, with the tracer obtained through the $^{195m}$Hg-$^{195m}$Au generator \cite{41}. Recently it is produced only in few laboratories \cite{42}. By using light charged particle beams in principle $^{195m}$Hg(T$_{1/2}$ = 41.6 h) can be NCA (non carrier-added) produced through the $^{197}$Au(p,3n), $^{197}$Au(d,4n), $^{nat}$Pt($\alpha$,xn) and $^{nat}$Pt($^3$He,xn) reactions (see Figures 16-19). 
In all cases $^{197g}$Hg (T$_{1/2}$ = 64.1 h) is simultaneously produced, except when using the $^{192}$Pt($\alpha$,n) or $^{194}$Pt($^3$He,2n) reactions, requiring enriched  targets. Concerning the production yields, the (d,4n) reaction is comparable to the (p,3n) reaction. The deuteron induced reactions have somewhat lower cross sections, the production requires higher energy but thinner targets. Alphas have lower cross sections, $^3$He induced reactions have much lower cross sections and the $^3$He gas is very expensive.

\begin{figure}
\includegraphics[scale=0.3]{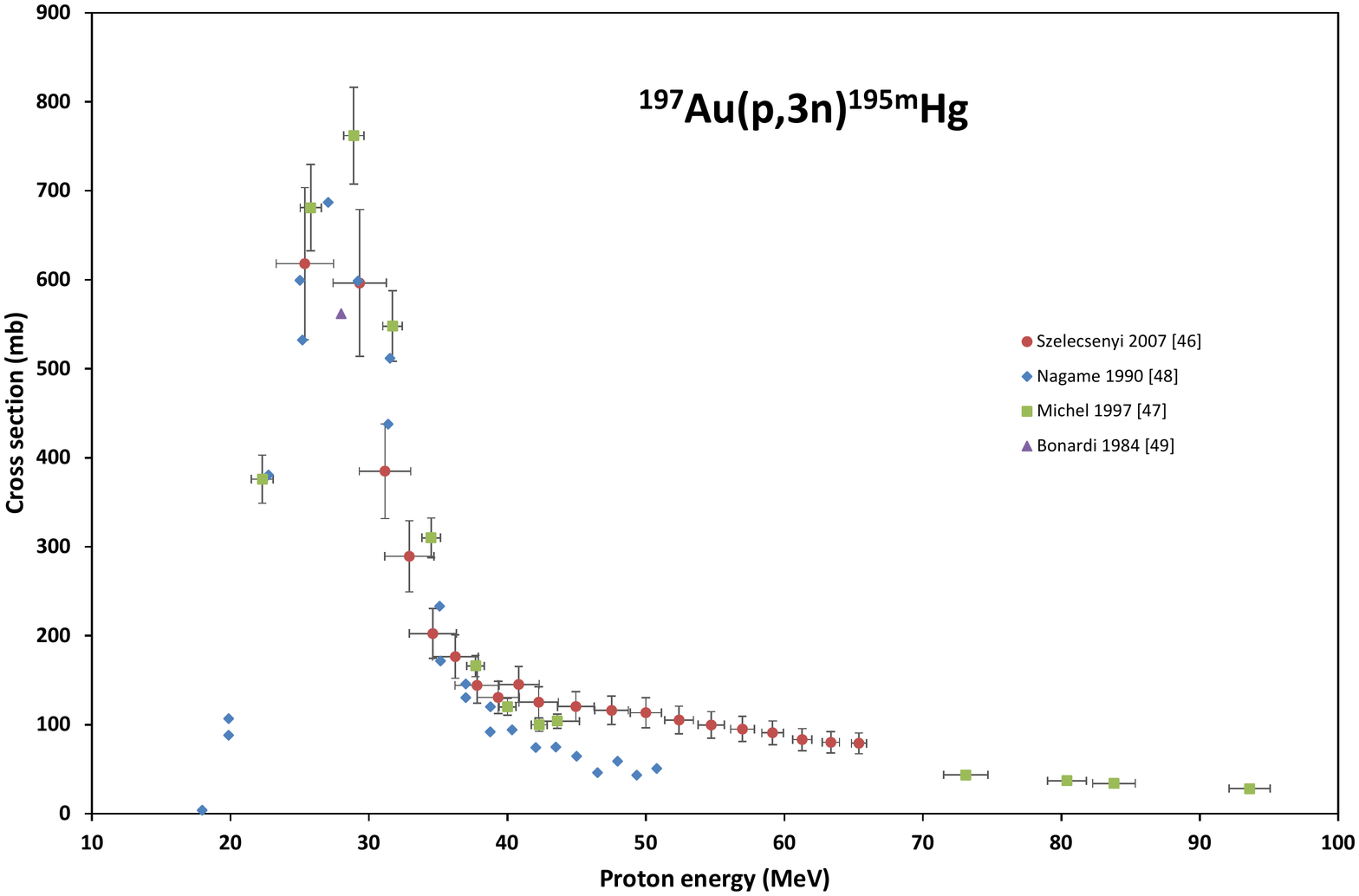}
\caption{16.	Measured cross-sections of the $^{197}$Au(p,3n)$^{195m}$Hg reaction (EXFOR) \cite{46,47,48,49}}
\end{figure}

\begin{figure}
\includegraphics[scale=0.3]{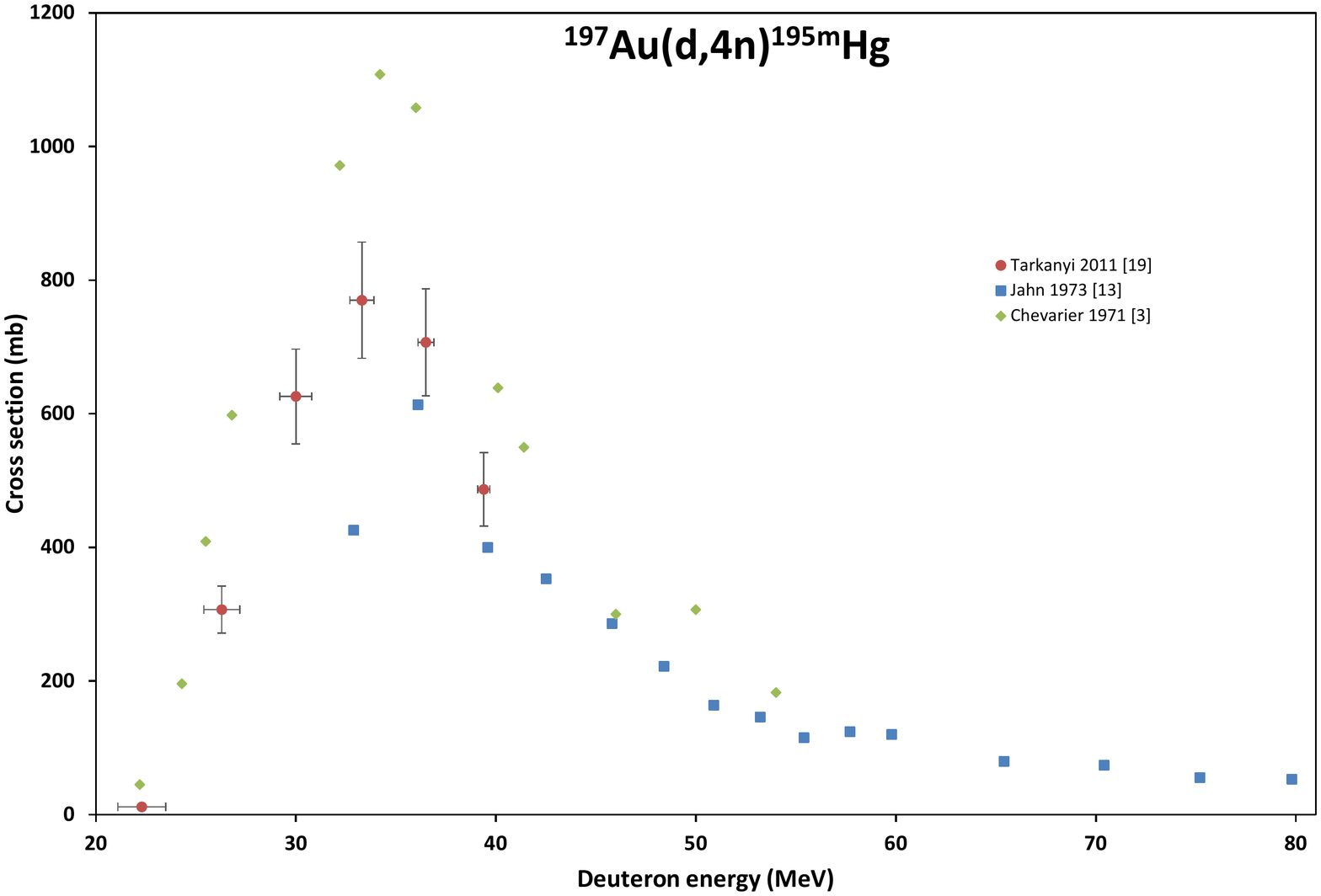}
\caption{Measured cross-sections of the $^{197}$Au(d,4n)$^{195m}$Hg reaction (EXFOR) \cite{3, 13, 19}}
\end{figure}

\begin{figure}
\includegraphics[scale=0.3]{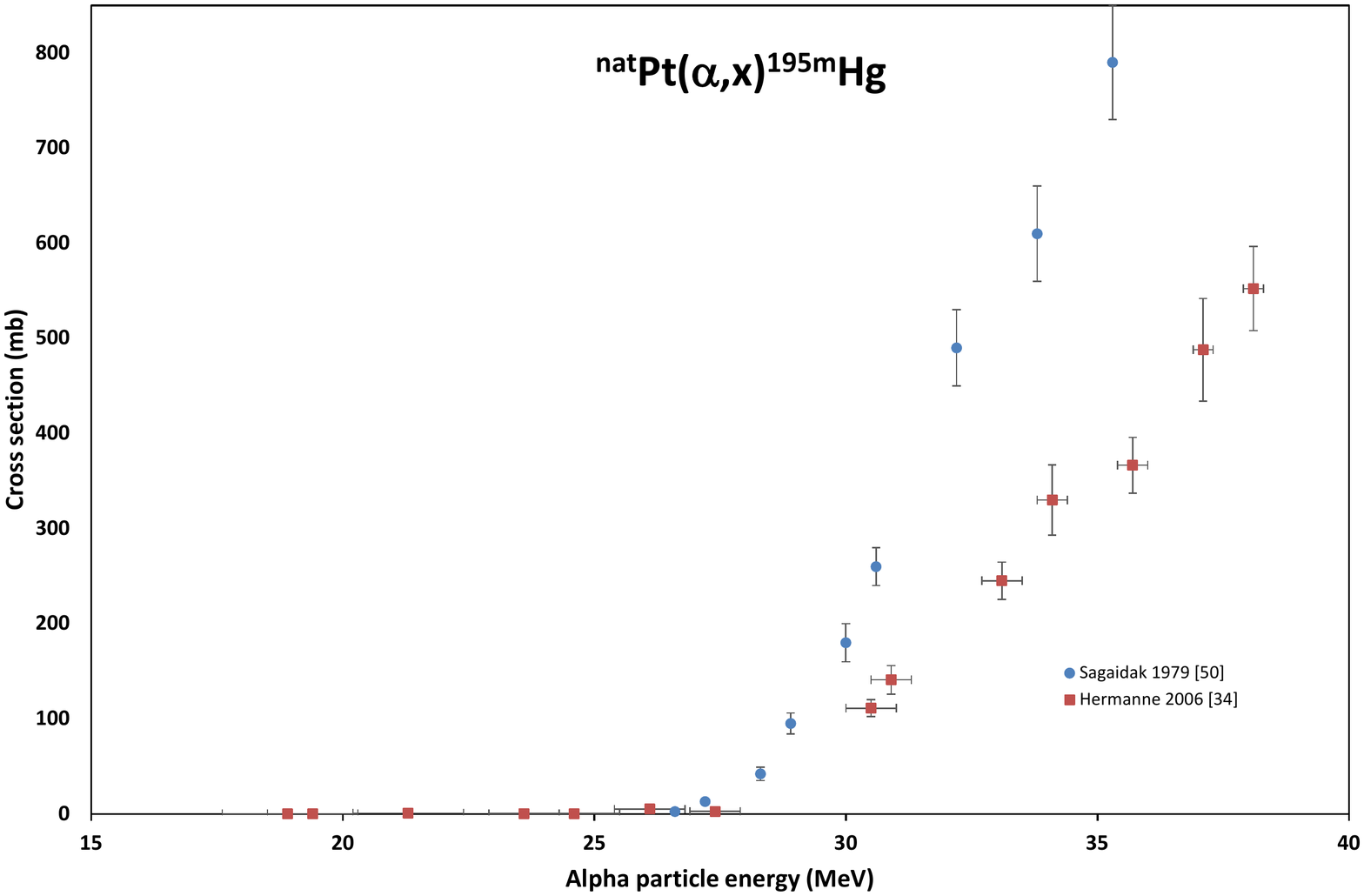}
\caption{Measured cross-sections of the $^{nat}$Pt($\alpha$,x)$^{195m}$Hg reaction (EXFOR) \cite{34, 50}}
\end{figure}

\begin{figure}
\includegraphics[scale=0.3]{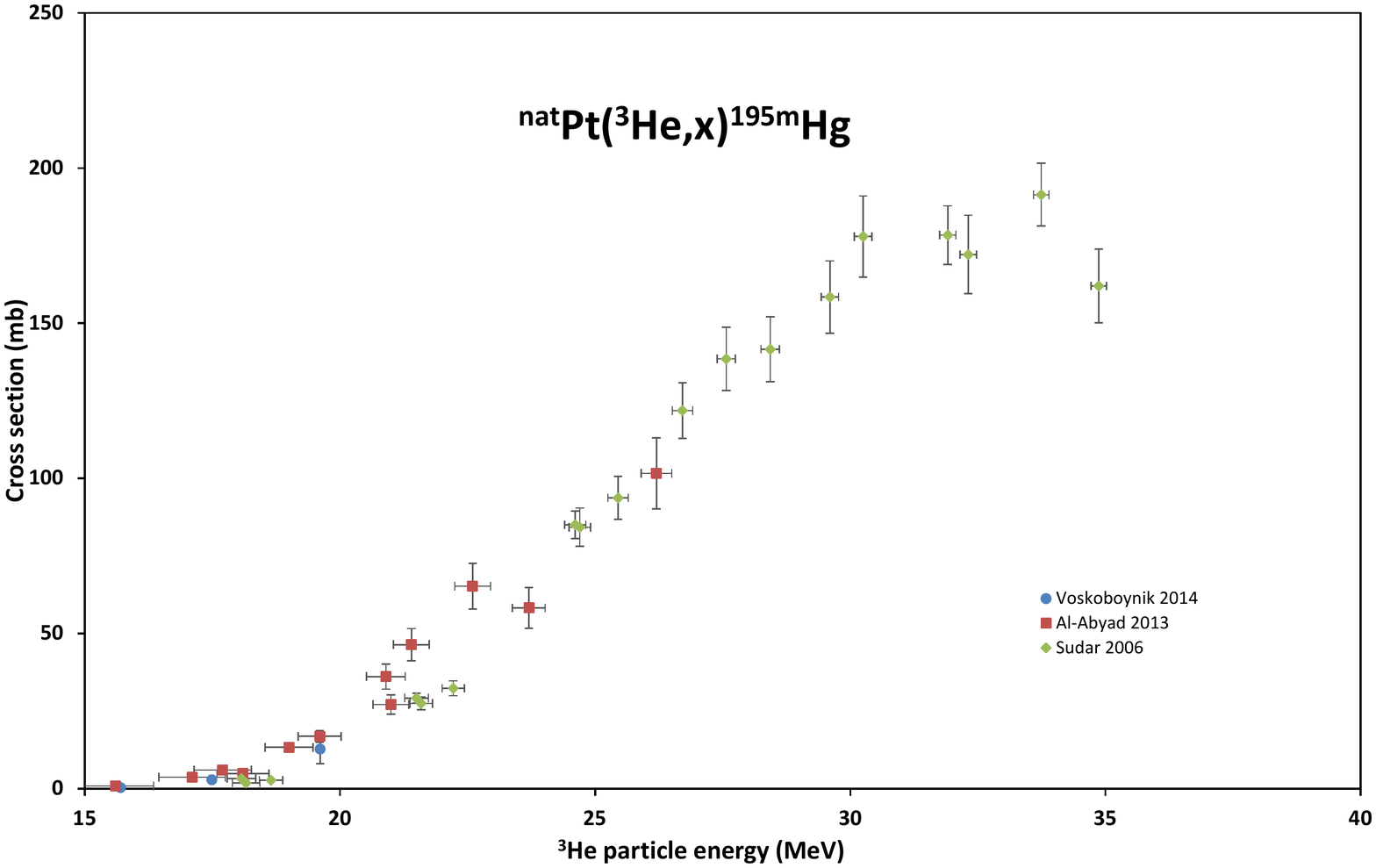}
\caption{Measured cross-sections of the $^{nat}$Pt($^3$He,x)$^{195m}$Hg reaction (EXFOR) \cite{35, 51, 52}}
\end{figure}

\subsection{$^{195g}$Au}
\label{6.2}
The long-lived $^{195g}$Au was proposed for targeted radiotherapy to kill single cells and small clusters \cite{33}. The production of $^{195g}$Au is feasible using many reactions. Three routes to get sufficiently pure $^{195}$Au (T$_{1/2}$ = 186.1 d) are used:  the $^{194}$Pt(d,n), $^{195}$Pt(p,n) or $^{193}$Ir($^3$He,n) reactions on highly enriched targets and appropriate limitation of incident particle energy  and/or applying long cooling times to let the shorter-lived Au by-products decay. No carrier added product with lower specific activity, because of the presence of isotopic contaminants, can be obtained from  targets with natural composition using the $^{nat}$Pt(p,xn), $^{nat}$Pt(d,xn), $^{nat}$Ir($\alpha$,xn), $^{nat}$Ir($^3$He,xn) routes. With bombardment of Au targets, ($^{nat}$Au(p,xn) and $^{nat}$Au(d,xn) reactions)  the product is carrier added. Even the $^{197}$Au($\alpha$,x)$^{195}$Au reaction was proposed for production \cite{43}. See figures 12, 20-22. In cases of $^{nat}$Ir+$\alpha$ and $^{nat}$Ir+$^3$He there are no data available in EXFOR.

\begin{figure}
\includegraphics[scale=0.3]{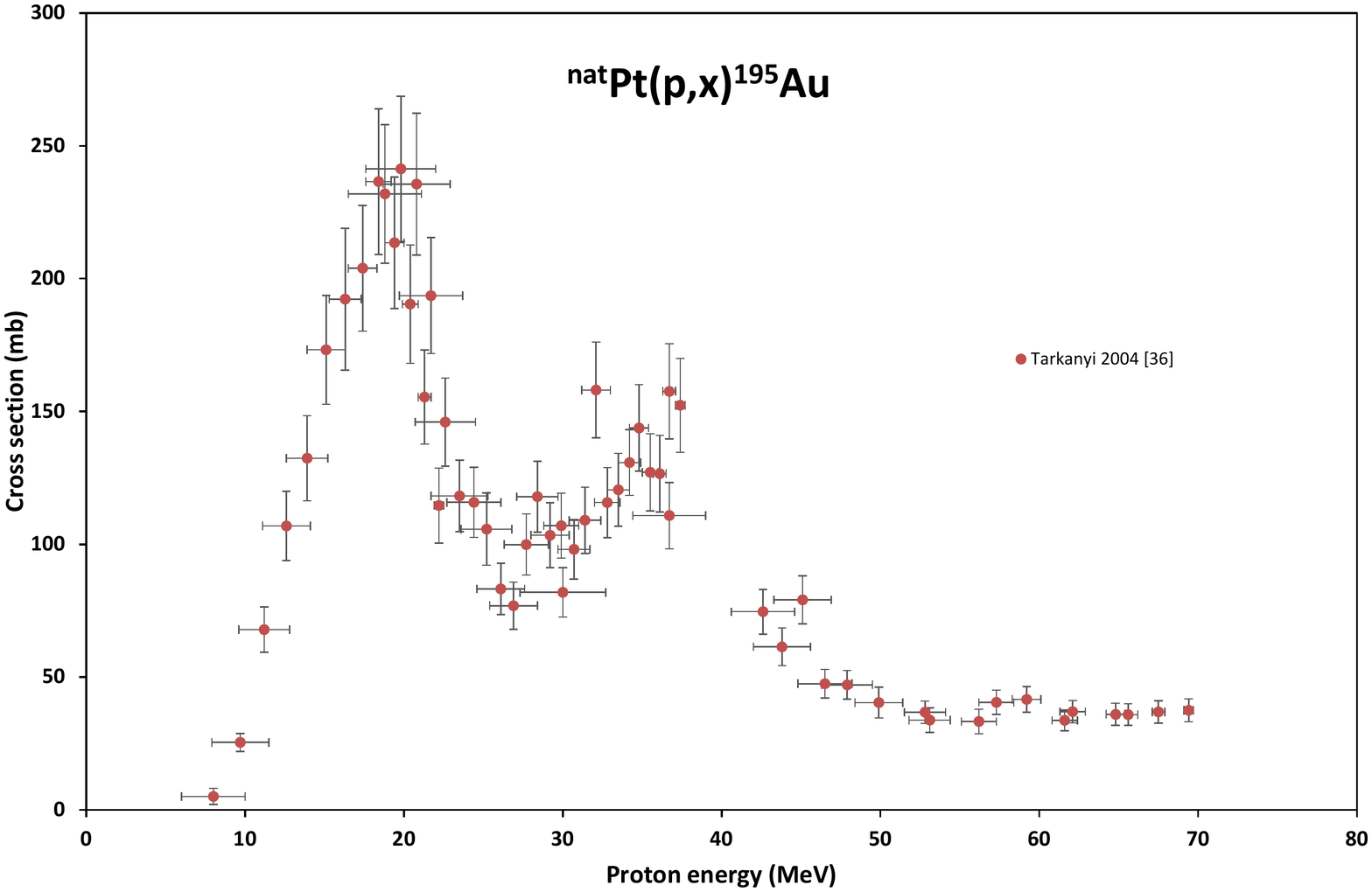}
\caption{Measured cross-sections of the $^{nat}$Pt(p,x)$^{195}$Au reaction (EXFOR) \cite{36}}
\end{figure}

\begin{figure}
\includegraphics[scale=0.3]{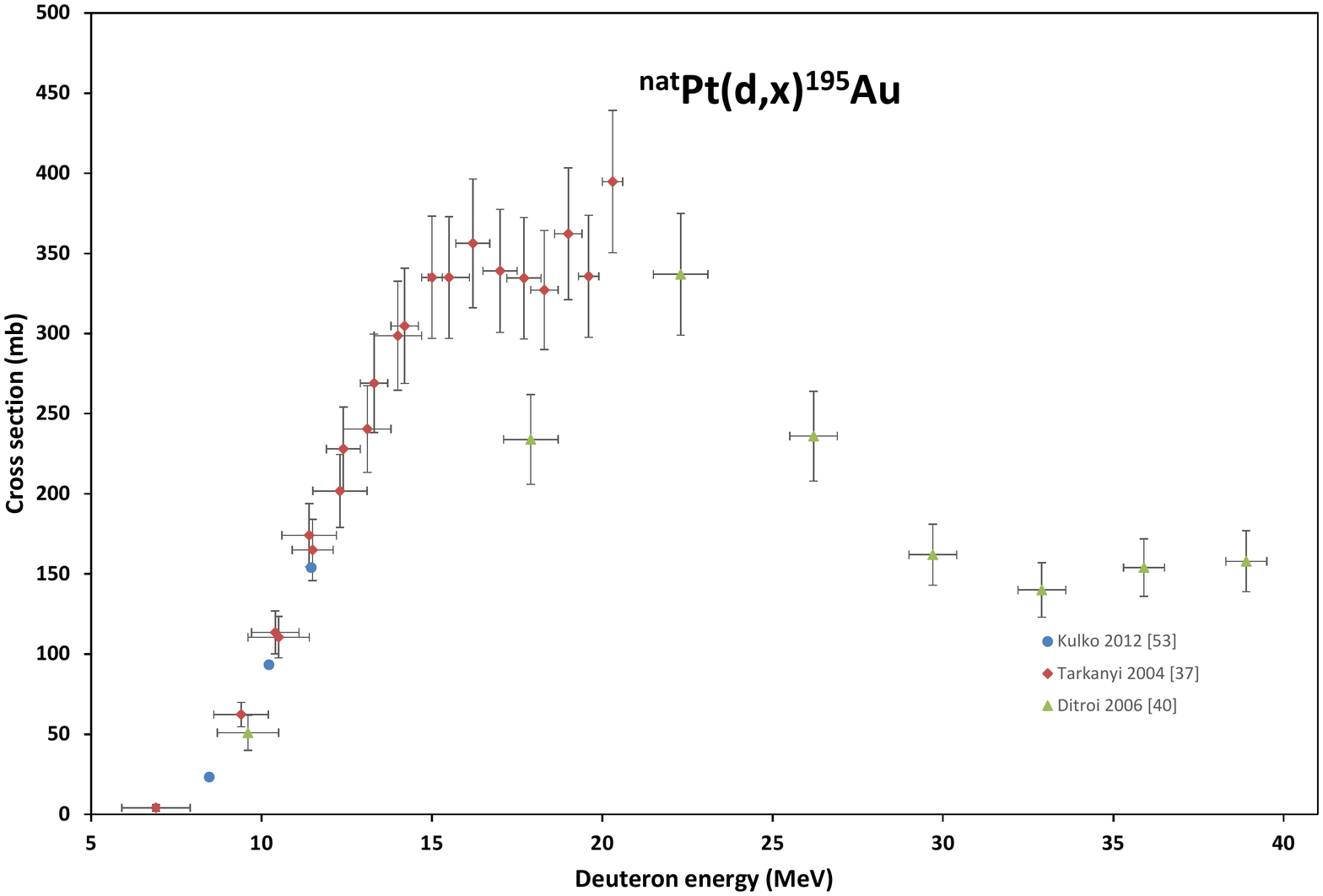}
\caption{Measured cross-sections of the $^{nat}$Pt(d,x)$^{195}$Au reaction (EXFOR) \cite{37, 40, 53}}
\end{figure}

\begin{figure}
\includegraphics[scale=0.3]{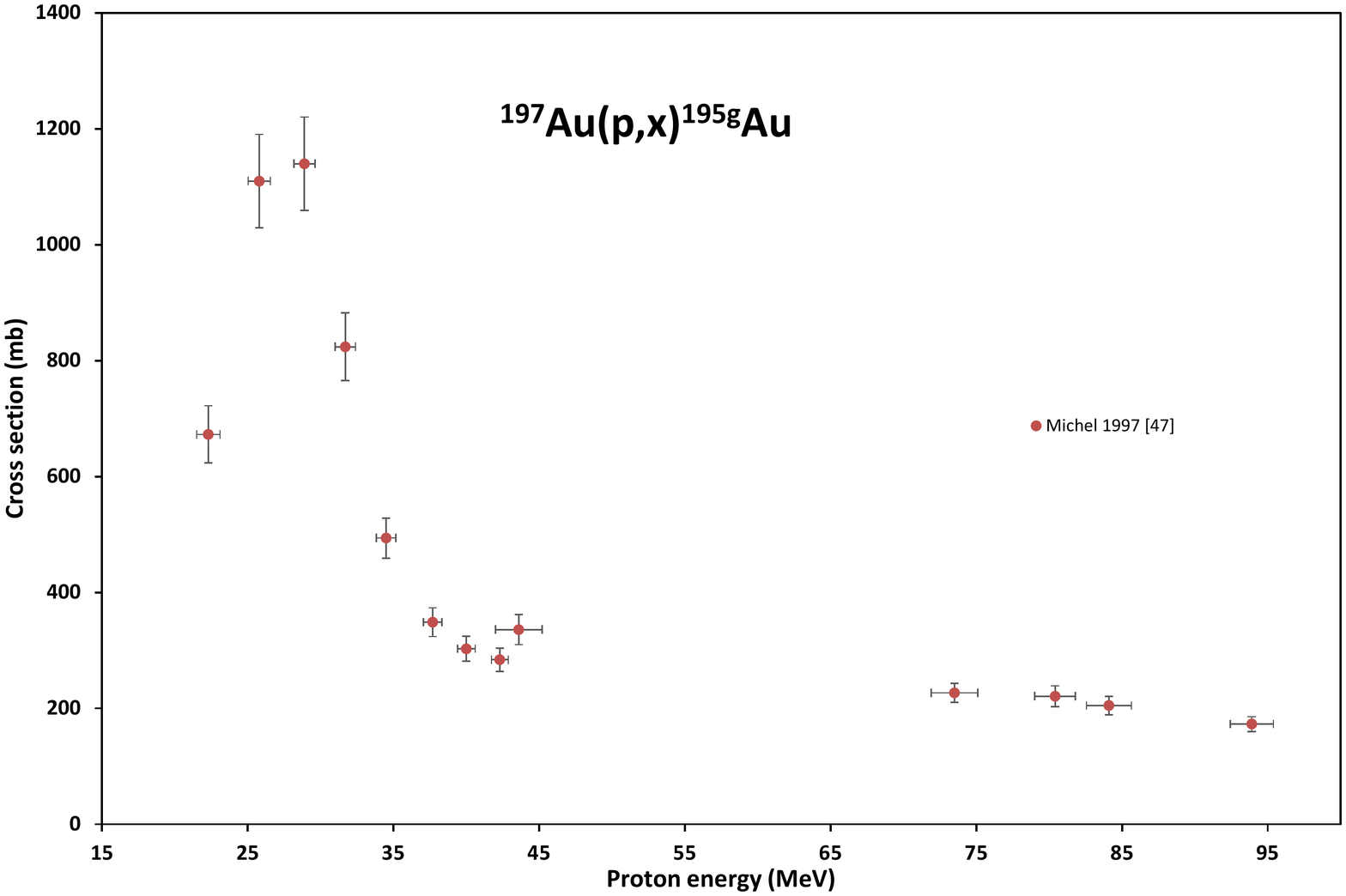}
\caption{Measured cross-sections of the $^{197}$Au(p,x)$^{195g}$Au reaction (EXFOR) \cite{47}}
\end{figure}

\subsection{$^{197m}$Hg and $^{197g}$Hg}
\label{6.3}
The radionuclides $^{197m}$Hg (T$_{1/2}$ = 23.8 h, E$_\gamma$ 133.98 keV, 33.5\%) and $^{197g}$Hg (T$_{1/2}$ = 64.14 h, E$_\gamma$ 77.4 keV, 18.7\%) are gamma-emitting radionuclides suitable for SPECT imaging and are of additional interest because of the potential therapeutic use at cellular level for their Auger- and conversion electron emission \cite{44}. They can be produced in a  no carrier added way  by using the $^{197}$Au(p,n), $^{197}$Au(d,2n), $^{195}$Pt($\alpha$,2n) and $^{196}$Pt($^3$He,2n) nuclear reactions. Natural Pt targets can also be used in alpha or 3He bombardment, but longer lived byproducts will be present. In the carrier added mode and with low specific activity the $^{196}$Hg(n,$\gamma$) \cite{45}, $^{nat}$Hg(n,$\gamma$), $^{nat}$Hg(p,x), $^{nat}$Hg(d,x) reactions  can be used. See figures 2, 3, 23-25. For the reactions $^{nat}$Hg+p and $^{nat}$Hg+d there are no data in EXFOR, but these reactions are under investigation by our group. 

\begin{figure}
\includegraphics[scale=0.3]{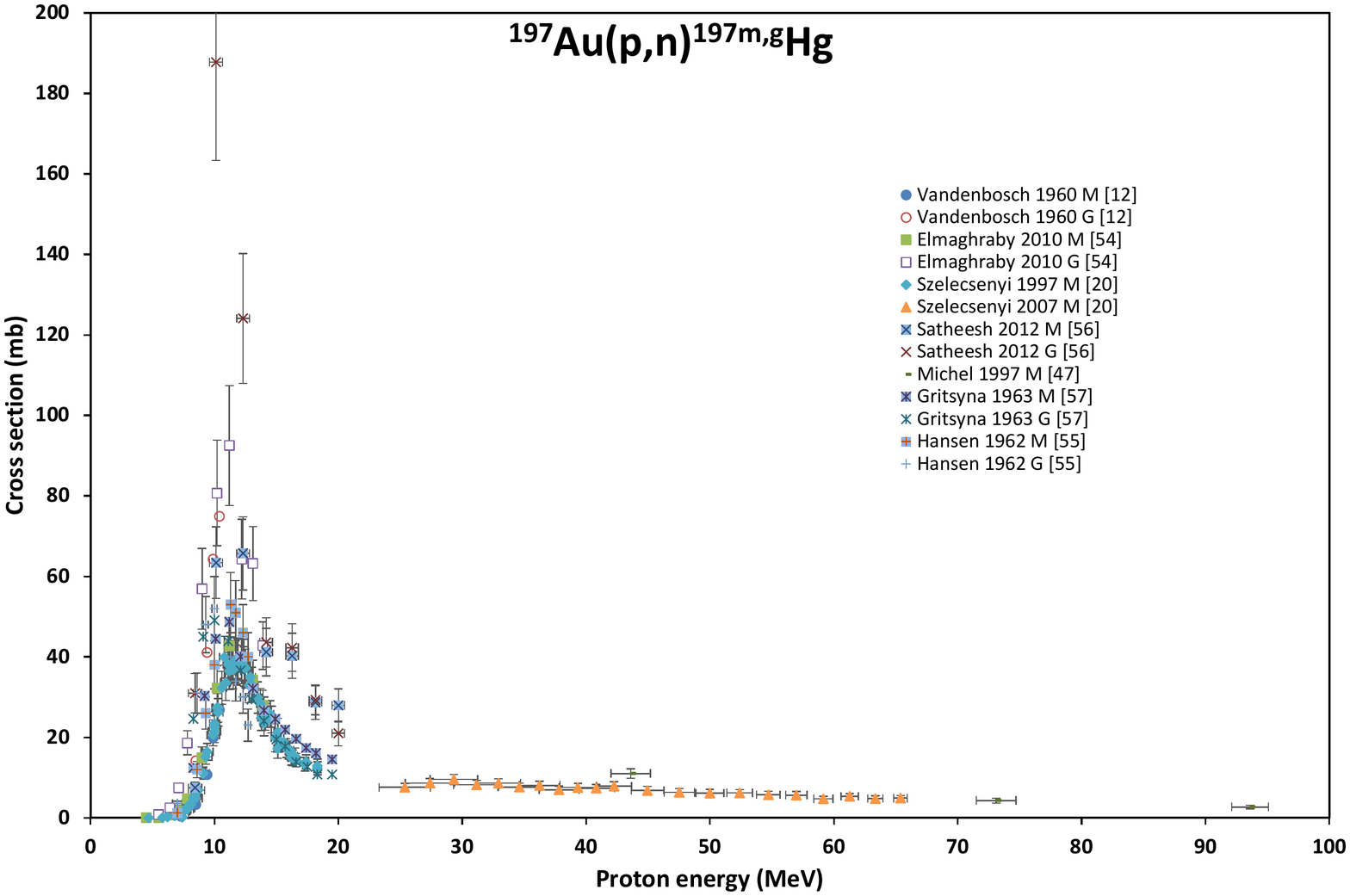}
\caption{Measured cross-sections of the $^{197}$Au(p,n)$^{197m}$Hg, $^{197g}$Hg reactions (EXFOR) \cite{12, 20, 46, 47, 54,55,56,57}}
\end{figure}

\begin{figure}
\includegraphics[scale=0.3]{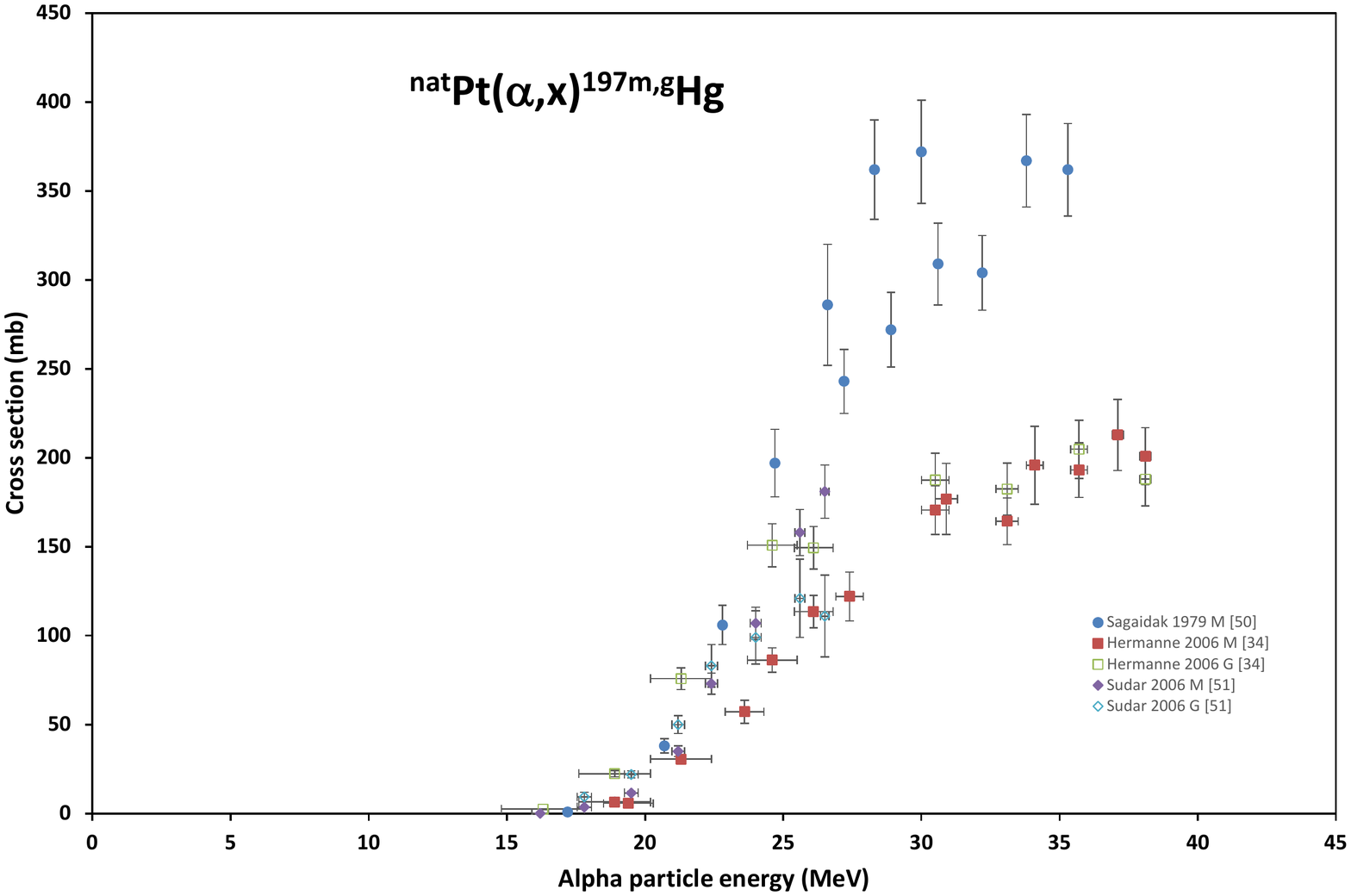}
\caption{Measured cross-sections of the $^{nat}$Pt($\alpha$,x)$^{197m}$Hg, $^{197g}$Hg reactions (EXFOR) \cite{34, 50, 51}}
\end{figure}

\begin{figure}
\includegraphics[scale=0.3]{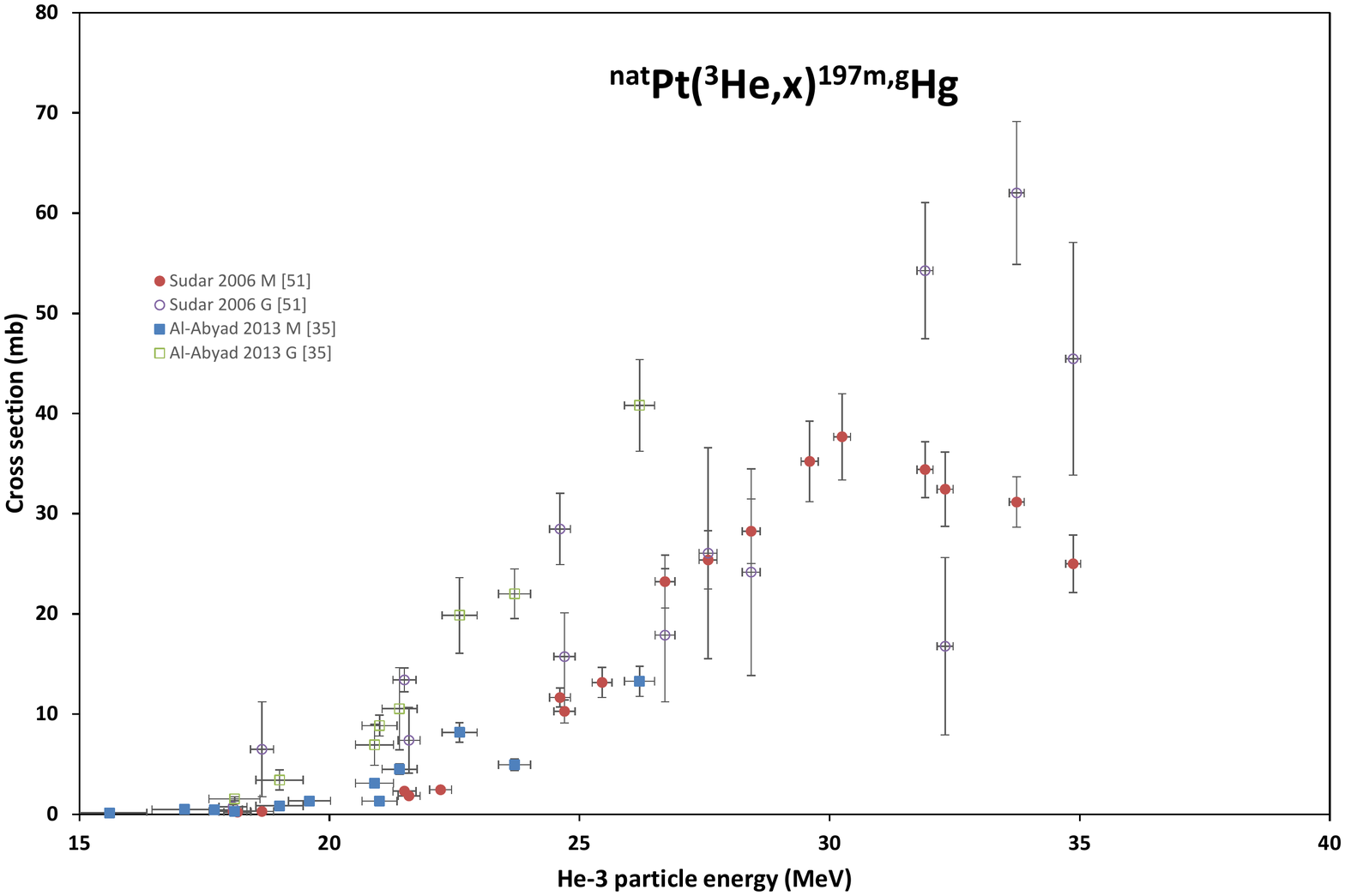}
\caption{Measured cross-sections of the $^{nat}$Pt($^3$He,x)$^{197m}$Hg, $^{197g}$Hg reactions (EXFOR) \cite{35, 51}}
\end{figure}

\subsection{$^{198g}$Au}
\label{6.4}
$^{198g}$Au (T$_{1/2}$ = 2.7 d) decays by 100\% $\beta^-$ and has been used for more than 40 years for tumor therapy and radionuclide synovectomy. Drawback is the emission of a 411 keV (96\%) $\gamma$-ray, which creates an unwanted radiation hazard for patients and personnel. $^{198}$Au is often produced, with low specific activity, at reactors via the $^{197}$Au(n,$\gamma$) reaction with a high thermal neutron capture cross-section. Irradiation of an enriched $^{198}$Pt target with protons or deuterons results in high specific activity, NCA $^{198}$Au . The presently investigated $^{197}$Au(d,p)$^{198g}$Au is similar to the $^{197}$Au(n,$\gamma$) process from the point of view of specific activity and results in a no-carrier-free product. Regarding to the Pt+p and Pt+d production routes the products are no carrier added, but the cross sections of the ground states are 2-3 times lower and the isomeric ratios  are significantly higher compared to $^{197}$Au(d,p) (see Figs. 26-27). No experimental data exist for Pt+p → $^{198m}$Au production). 

\begin{figure}
\includegraphics[scale=0.3]{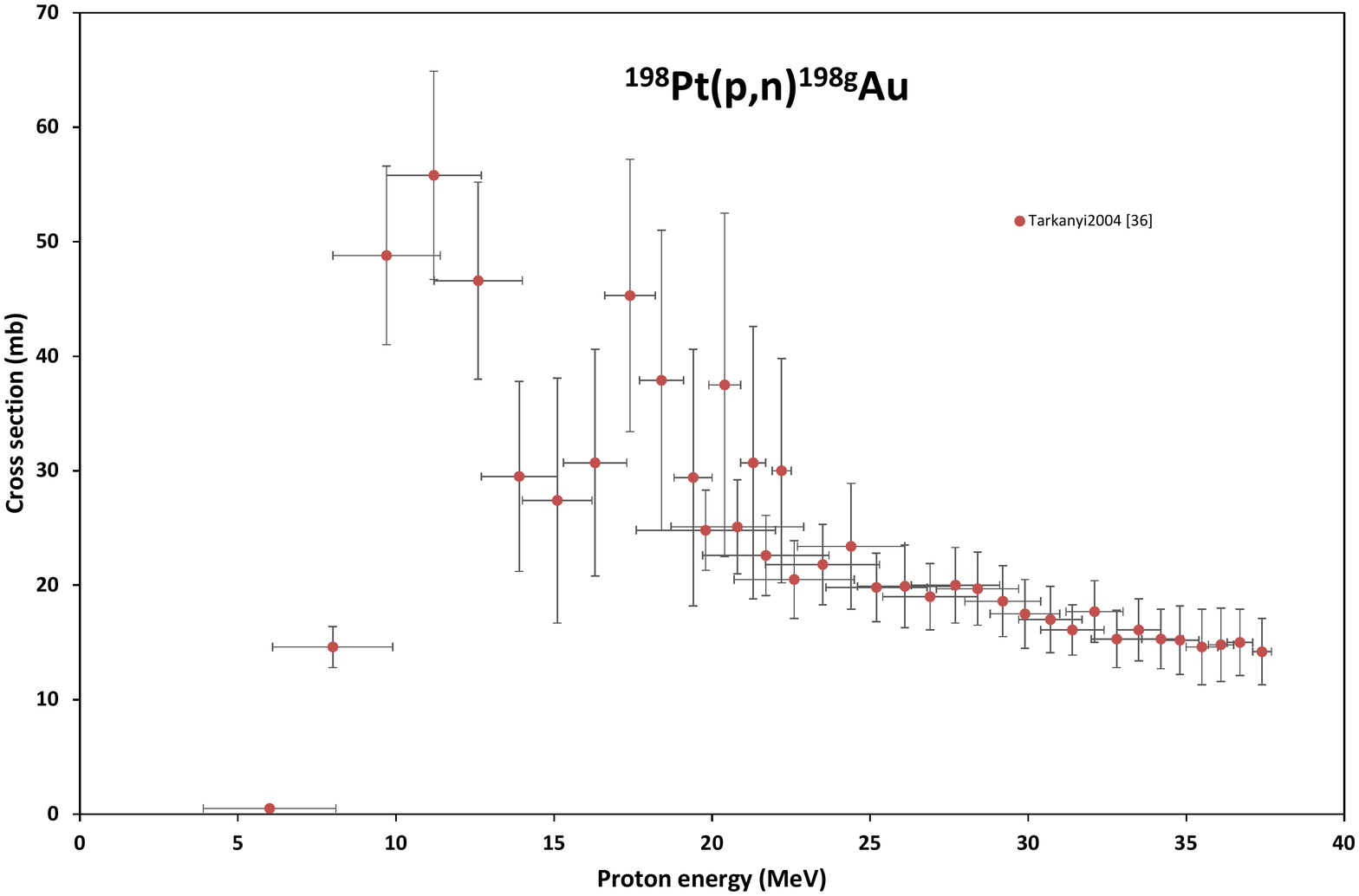}
\caption{Measured cross-sections of the $^{198}$Pt(p,n)$^{198g}$Au reaction (EXFOR) \cite{38}}
\end{figure}

\begin{figure}
\includegraphics[scale=0.3]{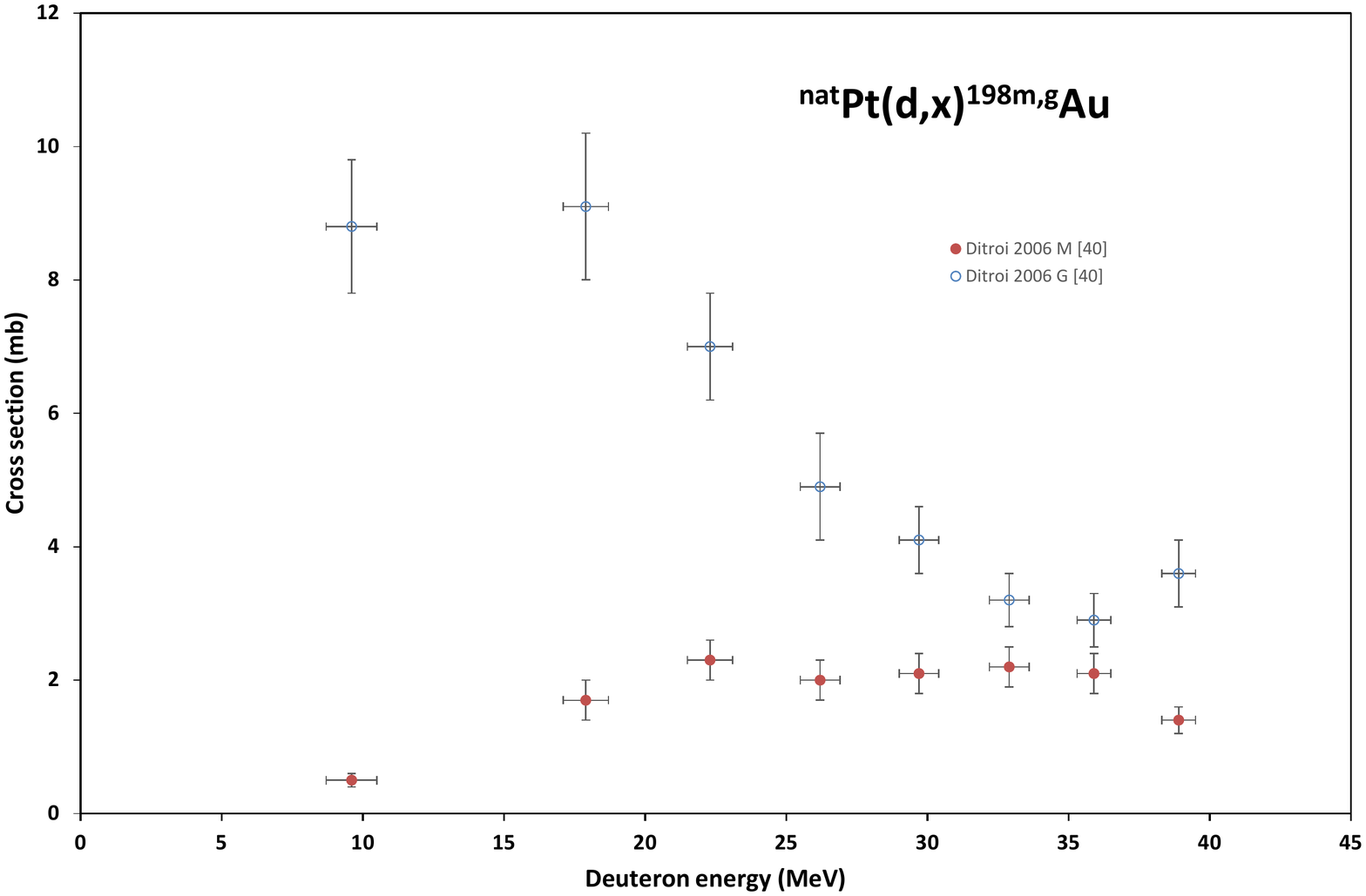}
\caption{Measured cross-sections of the $^{nat}$t(d,x)$^{198g}$Au reaction (EXFOR) \cite{40}}
\end{figure}

\section{Application for monitoring the deuteron beam parameters}
\label{7}

Gold is a frequently used as target backing material in several nuclear reaction cross section measurements and for routine isotope production due to the excellent chemical, mechanical and thermal properties.  Most cross section measurements are indirect, relative to a standard monitor reaction. Use of monitors is simpler and more convenient than direct beam intensity and beam energy measurement (no interference in any beam-target geometry). There are however strict requirements for the physical and chemical properties (the purity, the availability of the monitor material), for the activation conditions (excitation function, disturbing secondary reactions) and decay characteristics (half-life, gamma energy/intensity) of the used reaction product. Last and not least the beam monitoring requires reliable standard cross section data. The gold target, the possible reaction products induced by neutrons and light charged particles and their excitation functions fulfill all these requirements except for the quality of the cross sections in case of charged particles. There are relatively few experimental data on the excitation functions and the quality of the data still requires improvements. For the presently investigated deuteron induced reactions, candidate monitor reactions are available for a broad energy range: the $^{197m}$Hg, $^{197}$Hg (Q = -3.6 MeV, T$_{1/2}$ = 23.8h and 64.1 h), $^{195m}$Hg (Q = -19.3 MeV, T$_{1/2}$ = 41.6 h), $^{193m}$Hg (Q = -35.4 MeV, T$_{1/2}$ = 11. 8h), $^{196m}$Au, $^{196g}$Au(m+) (Q= -10.3 MeV, T$_{1/2}$ =  9.6 h and 6.2 d).

\section{Summary and conclusions}
\label{8}

The energy range for  excitation functions of the  $^{197}$Au(d,xn)$^{197m,197g,195m,195g}$Hg and $^{197}$Au(d,x)$^{198m,198g,196m,196g,194}$Au nuclear reactions that we studied earlier \cite{19} was extended up to 50 MeV. A good agreement was found in the overlapping energy regions when our new experimental values are compared with those published in our previous study. The agreement with the results of other authors can be considered also acceptable. The experimental data were compared with the results of the ALICE-D and EMPIRE-D codes and with TALYS data listed in different TENDL libraries. The agreement between code calculations and experimental data is lower in the case of deuterons than for proton and alpha particles induced reactions. In the code calculations, especially for description of isomers production, serious drawbacks appear: for the formation of Hg isotopes no changes are noted between the 2013 and 2014 versions; For Au radioisotopes some significant improvements of the prediction quality can be observed in the newest TENDL versions (2013-2014), but in some cases the 2014 predictions are even worse (c. f. $^{198g}$Au).
Use of deuteron induced reactions on Au for production of medically used $^{195m}$Au, $^{195g}$Au, $^{198g}$Au, $^{197m}$Hg and $^{197g}$Hg radioisotopes are discussed and compared to other production routes. From these comparisons we can conclude that:
\begin{itemize}
\item Some products ($^{195m}$Au, $^{195g}$Au) can be produced only via charged particle induced reactions
\item Charged particle reactions give possibility for high specific activity carrier free or no carrier added  products 
\item Investigation of the nuclear data shows that in this mass region, as it was proven earlier, the (d,2n) reaction has higher yield compared to the corresponding (p,n) reaction \cite{37}
\item	Additional products that cannot be formed with protons from the same element can be reached  by deuterons using the (d,p) reactions (i.e. $^{198}$Au)
\item	For local use, when only a small number of patients are treated and perhaps not regularly,  the radionuclides normally obtained via (n,$\gamma$), can be produced via (d,p) reactions.
\end{itemize}

\section{Acknowledgements}

This work was performed in the frame of the HAS-FWO Vlaanderen (Hungary-Belgium) project. The authors acknowledge the support of the research project and of the respective institutions (CYRIC, VUB, LLN) in providing the beam time and experimental facilities.
%\FloatBarrier
 


%% References

%\clearpage
\bibliographystyle{elsarticle-num}
\bibliography{Aud}






% \end{thebibliography}

\end{document}